\documentclass[usenatbib]{mn2e}
\usepackage{graphicx}

\title[Statistical properties of the combined emission of discrete
sources]{Statistical properties of the combined emission of
a population of discrete sources: astrophysical implications}

\author[M.Gilfanov et al.]
       {M.~Gilfanov$^{1,2}$, H.-J. Grimm$^1$, R.~Sunyaev$^{1,2}$\\
        $^1$Max-Planck-Institut f\"ur Astrophysik, 85741 Garching
            b. M\"unchen, Germany\\
	$^2$Space Research Institute, Moscow, Russia}
\date{\today}

\pubyear{2002}

\begin{document}

\maketitle

\begin{abstract}
We study the statistical properties of the combined emission of a
population of discrete sources 
(e.g.  X-ray emission of a  galaxy due to its X-ray binaries
population). Namely, we consider the  dependence of their total
luminosity $L_{\rm tot}=\sum L_k$ and of fractional $rms_{\,\rm
tot}$ of their variability on the number of sources $n$ or, equivalently,
on the normalization of the luminosity function. We show that due to
small number statistics a regime exists, in which $L_{\rm tot}$
grows non-linearly with $n$, in an apparent
contradiction with the seemingly obvious prediction    
$\left  <L_{\rm tot} \right > =
\int L\,\frac{dN}{dL}\, dL 
\propto n
$.
In this non-linear regime, the $rms_{\,\rm tot}$ decreases with $n$
significantly more slowly than expected from the 
$rms\propto 1/\sqrt{n}$ averaging law. For example, for a
power law luminosity function with a slope of $\alpha=3/2$, in the
non-linear regime, $L_{\rm tot}\propto n^2$ and 
the $rms_{\,\rm tot}$ does not depend at all on the  number of
sources $n$. Only in the limit of $n\rightarrow\infty$ do these
quantities behave as intuitively expected, 
$L_{\rm tot}\propto n$ and $rms_{\,\rm tot}\propto 1/\sqrt{n}$.
We give exact solutions  and derive
convenient analytical approximations for  $L_{\rm tot}$ and
$rms_{\,\rm tot}$.   

Using the total X-ray luminosity of a galaxy due to its X-ray binary
population as an example, we show that the $L_X-$SFR and $L_X-M_*$ relations
predicted from the respective ``universal'' luminosity functions of
high and low mass X-ray binaries are in a good agreement with
observations. Although  caused by small number statistics the
non-linear regime in these examples extends as far as  
SFR$\la 4-5$ M$_\odot$/yr and $\log(M_*/M_\odot)\la 10.0-10.5$,
respectively.

\end{abstract}
\begin{keywords}
methods: statistical -- methods: data analysis --
X-rays: galaxies -- X-rays: binaries

\end{keywords}

\section{Introduction and qualitative consideration}
\label{sec:intro}

In many astrophysical situations a problem arises to predict or
interpret results of measurements of the total (combined) luminosity of
a population of discrete sources. 
Among many examples are the total luminosity of X-ray 
binaries in a galaxy, or total flux of background sources
detected above the sensitivity limit inside the field of view of a
telescope.

In the following discussion we will use  high mass X-ray binaries
(HMXB) in star forming galaxies as an example.
As was shown by \citet*{grimm}, the luminosity distribution of HMXB
sources in a galaxy is described to first approximation by a
''universal'' luminosity function whose shape is the same in all
galaxies and whose normalization is proportional to the star formation
rate (SFR) of the parent galaxy:
\begin{eqnarray}
\frac{dN}{dL}\propto{\rm SFR}\times f(L)
\label{eq:xlf}
\end{eqnarray}
In a broad luminosity range, $\log(L_X)\sim 35.5-40.5$, the shape of the
HMXB ``universal'' luminosity function is close to a power 
law, $f(L)=L^{-\alpha}$, with the slope of $\alpha\approx 1.6$.
Importantly, the luminosity of the compact sources in star forming
galaxies appears to be restricted by a maximum value of  
$L_{\rm cut}\sim {\rm few}\times 10^{40}$ erg/s. This cut-off
luminosity can be defined, for example, by the Eddington luminosity
limit for the most massive objects associated with the star forming
regions. Obviously, on the faint side, the luminosity
distribution eq.(\ref{eq:xlf}) must become flatter or have a cut-off
as well, in order to keep the total number of sources finite. This low
luminosity cut-off may be caused
for example by the propeller effect \citep{propeller} as discussed by
\citet{lmc}.  

The expectation value for the total number of sources in
a galaxy equals  
\begin{eqnarray}
\left< N_{\rm tot} \right> =\int_0^{+\infty} \frac{dN}{dL} dL
\propto {\rm SFR}
\label{eq:nmean}
\end{eqnarray}
and, naturally, is directly proportional to its star formation
rate. The number of sources actually observed in a given galaxy obeys
Poisson distribution $p_{\mu}(N)$ with $\mu$ defined by 
eq.(\ref{eq:nmean}). 
Apart from effects of counting statistics, the number of HMXB
sources found in an arbitrarily chosen galaxy will be close to 
the above expectation value.

The problem, considered in this paper, is the behavior of the total 
luminosity of high mass X-ray binaries in a galaxy
\begin{eqnarray}
L_{\rm tot}=\sum_{k=1}^{k=N} L_k
\label{eq:ltot_sum}
\end{eqnarray}
as a function of its star formation rate.

An apparently obvious expression for the $L_{\rm tot}$ can be
obtained integrating the luminosity distribution (\ref{eq:xlf}):
\begin{eqnarray}
\left  <L_{\rm tot} \right > =
\int_0^{+\infty} L\,\frac{dN}{dL}\, dL 
\propto {\rm SFR}
\label{eq:ltot_mean}
\end{eqnarray}
Hence, one might expect that the total X-ray luminosity of HMXB
sources is also directly proportional to the star formation rate of
the host galaxy, as is the total number of sources. 
This problem, however, involves some subtleties
related to the statistical properties of the power law distribution of
the sources over luminosity, which  appear not to have been recognized
previously, at least in astrophysical context (a somewhat related
problem has been considered by \citealt{kalogera} in connection with
estimating the coalescence rate for the NS--NS binaries in the Galaxy).  
Although eq.(\ref{eq:ltot_mean}) correctly predicts 
{\em the average X-ray luminosity computed for a large number of
galaxies} with similar values of star formation rate, it fails to
describe the relation between {\em the most probable value of X-ray
luminosity of an arbitrarily chosen galaxy} and its SFR.
The main surprise of the study presented here is that, in the low SFR
regime, the relation between SFR of the host galaxy and the total
luminosity of its HMXBs is non-linear -- with increase of the
star formation rate the luminosity appears to grow faster than
linear. The relation becomes linear only for sufficiently high
star formation rates, when the total number of objects with a
luminosity close to the maximum possible value, defined by 
$L_{\rm cut}$, becomes sufficiently large.

This can be illustrated by the following simple consideration. 
For an arbitrarily chosen galaxy, the brightest source is most
likely going to have a luminosity $\tilde{L}_{\rm max}$ defined by the 
condition\footnote{Indeed, for example sources 10 times more
luminous will appear on average in one out of $\sim10^{(\alpha-1)}$
galaxies.}
\begin{eqnarray}
N(L>\tilde{L}_{\rm max})=\int_{\tilde{L}_{\rm max}}^{+\infty} 
\frac{dN}{dL}\, dL \sim 1
\end{eqnarray}
For a power law luminosity distribution with slope $\alpha$ and with a
cut-off at $L_{\rm cut}$, eq.(\ref{eq:xlf}), the above expression
yields:
\begin{eqnarray}
\renewcommand{\arraystretch}{1.5}
\begin{array}{lll}
\tilde{L}_{\rm max}&  \propto SFR^{\ \frac{1}{\alpha-1}}  & \mbox{low SFR}\\
\tilde{L}_{\rm max}& =L_{\rm cut}	            & \mbox{high SFR}
\end{array}
\end{eqnarray}

As might be intuitively expected, at low SFR, the most probable
luminosity of the brightest source increases with SFR, until it
reaches the maximum value of $L_{\rm cut}$. The threshold
value of the star formation rate, separating low and high SFR regimes,
is defined by the condition $N(L\sim L_{\rm cut})\sim 1$,
i.e. that there are $\sim$few sources expected, with luminosity close
to the cut-off value $L_{\rm cut}$.  

The most probable value of the total luminosity, $\tilde{L}_{\rm tot}$
can be then computed  integrating the luminosity function from 
$L_{min}$ to $\tilde{L}_{\rm max}$:
\begin{eqnarray}
\tilde{L}_{\rm tot} \approx \int^{\tilde{L}_{\rm max}}_{L_{min}} \frac{dN}{dL}
L~dL 
\end{eqnarray}
which, for $1< \alpha < 2$ and $ L_{\rm min} << \tilde{L}_{\rm max}$,
leads to 
\begin{eqnarray}
\renewcommand{\arraystretch}{1.5}
\tilde{L}_{\rm tot} \propto \left\{ \begin{array}{ll}
SFR^{\frac{1}{\alpha-1}}& \mbox{low SFR}\\
SFR	& \mbox{high SFR}
\end{array}
\right.
\label{eq:ltot_intro}
\end{eqnarray}
i.e. is non-linear in the low SFR regime and becomes linear only at
high star formation rates. 

This can be qualitatively understood as follows. For a slope of the
luminosity distribution $1< \alpha < 2$, the total luminosity of a
galaxy is defined by the brightest sources. The non-linear behavior in
the low-SFR limit is caused by the fact that an increase of the SFR
leads to non-linear increase of the luminosity of the brightest
sources. Therefore their total luminosity grows faster 
than the star formation rate. This non-linear growth continues until
the maximum possible value of the luminosity of the compact sources is
achieved. Further increase of the star formation rate leads to
a linear increase of the number of the brightest sources in the galaxy,
but not of their individual luminosities. Consequently, the
$L_X-$SFR relation becomes linear.

In the more formal language of statistics such a behavior is related
to the properties of the probability distribution of the collective
luminosity $p\,(L_{\rm tot})$. In particular, it can be understood in
terms of the difference between the expectation mean and the mode of
the probability distribution. The expectation mean is defined as 
\begin{eqnarray}
\left<L_{\rm tot}\right>=\int_0^\infty p\,(L_{\rm tot})
\,L_{\rm tot}\,dL_{\rm tot}
\label{eq:ltot_mean_2}
\end{eqnarray}
and is given by eq.(\ref{eq:ltot_mean}).  
The mode of the statistical distribution, $\tilde{L}_{\rm tot}$, is
defined as the value of the random variable ($L_{\rm tot}$ in our
case) at which the probability distribution $p\,(L_{\rm tot})$
reaches its maximum value. 
Whereas the expectation mean $\left<L_{\rm tot}\right>$ describes the
result of averaging of the X-ray luminosities of many galaxies having
similar values of SFR, it is the mode of the  $p\,(L_{\rm tot})$
distribution that predicts the most probable value of the total
luminosity of a randomly chosen galaxy.
The non-linear behavior in the low SFR regime is caused by the  
skewness of the probability distribution $p\,(L_{\rm tot})$ resulting
in a difference between expectation mean and mode. 
In the high SFR limit, the $p\,(L_{\rm tot})$ distribution 
asymptotically approaches the Gaussian distribution, in accord with 
the Central Limit Theorem. 
The boundary value of SFR, separating the non-linear and linear
regimes of the $L_X-$SFR relation is defined by the parameters of the
luminosity function. 

Interestingly, the fact of the existence of a linear regime in the
$L_X$--SFR relation is a direct consequence of the 
cut-off in the luminosity function. Only in the presence of a
maximum possible luminosity for the sources, $L_{\rm cut}$, (for
instance Eddington limit for a neutron star) a linear regime can 
exist, when the total luminosity of a galaxy is defined by a
sufficiently large number of bright sources near $L_{\rm cut}$, and
any subsequent increase of the star formation rate results in
linear growth of the total luminosity.

In the above discussion, we used high mass X-ray binary
populations in star forming galaxies as an example.
Obviously, the effect considered in this paper is of a broader
general interest and is at work in many situations related 
to computing/measuring integrated luminosity of a finite number of 
discrete sources. One example related to the application of
stellar synthesis models to observations of stellar clusters has been
independently discovered by \citet{cervino04}.

The paper is structured as follows. In section \ref{sec:ltot_section}
we consider the statistical properties of the total luminosity, in
particular in section \ref{sec:probdist} derive the formulae for the
probability  distribution $p\,(L_{\rm tot})$, using two different
approaches, and present results of numerical calculations in
\ref{sec:examples}. The variability of the total emission is studied
in section \ref{sec:variability}. In section \ref{sec:applications} we
discuss astrophysical applications, including the properties of the
total X-ray emission of a galaxy due to its population of high-
(section \ref{sec:hmxb}, \ref{sec:imbh}) and low- (section
\ref{sec:lmxb}) mass X-ray binaries. Our results are summarized in
section \ref{sec:summary}. In the Appendixes we derive convenient
approximations for $\tilde{L}_{\rm tot}$ and fractional $rms$ of the
total emission and consider their asymptotical behaviour. A casual
reader, not interested in the mathematical aspects of the problem, can
skip section \ref{sec:probdist} and proceed with section
\ref{sec:examples}.

\section{Total luminosity}
\label{sec:ltot_section}

We consider a population of sources with a luminosity function (LF):
\begin{eqnarray}
\frac{dN}{dL}=A\,f(L)
\label{eq:lfd}
\end{eqnarray}
The expectation values for the total number of sources and  
the total luminosity are:  
\begin{eqnarray}
\left< n\right>=\int_{0}^{\infty} \frac{dN}{dL} \, dL
\label{eq:ntot}
\end{eqnarray}
\begin{eqnarray}
\left< L_{\rm tot}\right> =\int_{0}^{\infty} L \, \frac{dN}{dL} \, dL
\label{eq:ltot}
\end{eqnarray}
It is assumed that both quantities are well defined and finite.

\subsection{The probability distribution of the total luminosity} 
\label{sec:probdist}

Below we present two methods of computing the probability
distribution of the total luminosity. The second of these two
methods is somewhat more convenient computationally.

\subsubsection{Method I}
\label{sec:method1}

To compute the probability distribution for the total luminosity,
$p\,(L_{\rm tot})$,  we divide the $L_1-L_2$ luminosity range into
intervals of infinitesimal width $\delta L_k$ and
express the combined luminosity of all sources as a sum
\begin{eqnarray}
L_{\rm tot}=\sum_k L_{\rm tot, k}
\label{eq:ltot_sum_m1}
\end{eqnarray}
where $L_{\rm tot, k}$ is the combined luminosity of the sources in
the $k$-th interval, running from $L_k$ to $L_k+\delta L_k$. 
The number of sources in the interval ($L_k$, $L_k+\delta L_k$) obeys
a Poisson distribution with mean 
\begin{eqnarray}
\mu_k=\frac{dN}{dL}(L=L_k)\, \delta L_k
\end{eqnarray}
For $\delta L_k\rightarrow 0$, it is sufficient to consider the
occurrence of either zero or one source per interval whose
probabilities are, respectively: 
\begin{eqnarray}
p_k(0)=1-\frac{dN}{dL}\, \delta L_k + O\left(\delta L_k^2\right)
\\ \nonumber
p_k(1)=\frac{dN}{dL}\, \delta L_k + O\left(\delta L_k^2\right)~~~~~
\end{eqnarray}
The probability distribution for the combined luminosity of the
sources in the $k$-th interval is 
\begin{eqnarray}
p\,(L_{\rm tot, k})=\delta(L_{\rm tot,k}) \left(1-\frac{dN}{dL}\, \delta
L_k\right)  + 
\nonumber
\\
\delta(L_{\rm tot,k}-L_k) \frac{dN}{dL}\, \delta L_k
\end{eqnarray}
where $\delta(x-x_0)$ is the delta function.
The characteristic function of this distribution is 
\begin{eqnarray}
\hat{p}\,(L_{\rm tot,k})=\int p\,(L_{\rm tot,k}) \, e^{i\omega L} \,
dL=
\nonumber
\\
1-\frac{dN}{dL}\, \delta L_k + \frac{dN}{dL}\, \delta L_k \,
e^{i\omega L_k}
\end{eqnarray}
Using the convolution theorem, the characteristic function of the
probability distribution of the total luminosity can be computed as a
product of characteristic functions of $p\,(L_{\rm tot, k})$
\begin{eqnarray}
\hat{p}\,(L_{\rm tot})=\prod_k \hat{p}\,(L_{\rm tot,k})
\label{eq:phat_ltot}
\end{eqnarray}
\begin{eqnarray}
\ln \hat{p}\,(L_{\rm tot})=\sum_k \ln(1-\frac{dN}{dL}\, \delta L_k +
\frac{dN}{dL}\, \delta L_k \, e^{i\omega L_k})=
\nonumber
\\
- \sum_k \frac{dN}{dL}\, \delta L_k + 
\sum_k \frac{dN}{dL}\, \delta L_k \, e^{i\omega L_k}
+O\left(\delta L_k^2\right)=
\nonumber
\\
-\int \frac{dN}{dL}\, dL+\int \frac{dN}{dL}\, e^{i\omega L}\, dL
\label{eq:ln_phat_ltot}
\end{eqnarray}
Finally,
\begin{eqnarray}
p\,(L_{\rm tot})=\int \hat{p}\,(L_{\rm tot}) \, 
e^{-i\omega L_{\rm tot}} \, d\omega =
\nonumber
\\
\label{eq:p_ltot}
\int exp\left(
\int \frac{dN}{dL}\, e^{i\omega L}\, dL  - \left< n \right> 
-i\omega L_{\rm tot} \right) \, d\omega
\end{eqnarray}
where $\left< n\right>$ is given by eq.(\ref{eq:ntot}).

\begin{figure*}
\centerline{
\vbox{
\hbox{
\resizebox{0.40\hsize}{!}{\includegraphics[bb=18 165 574 705]{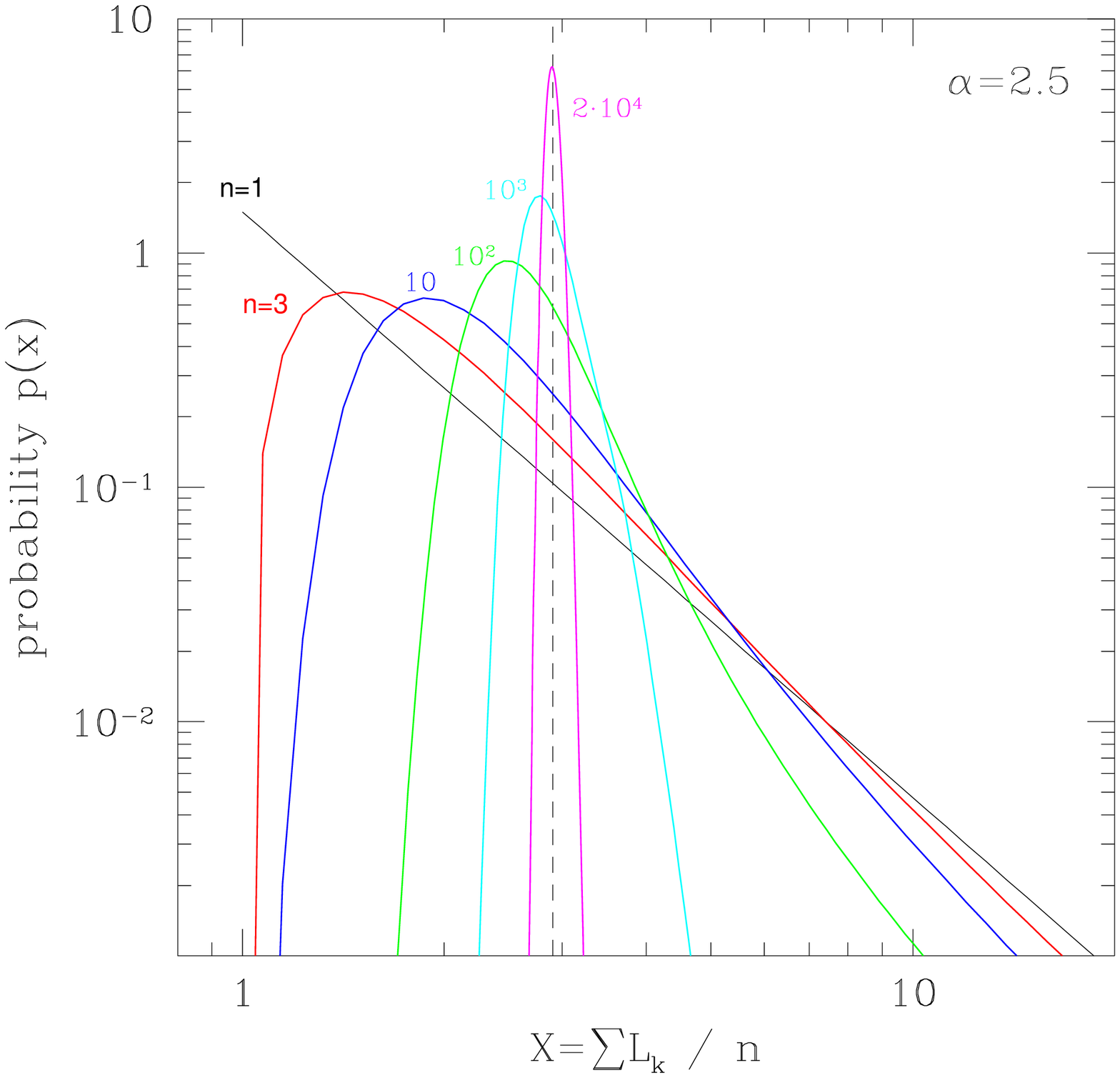}}
\resizebox{0.40\hsize}{!}{\includegraphics[bb=18 165 574 705]{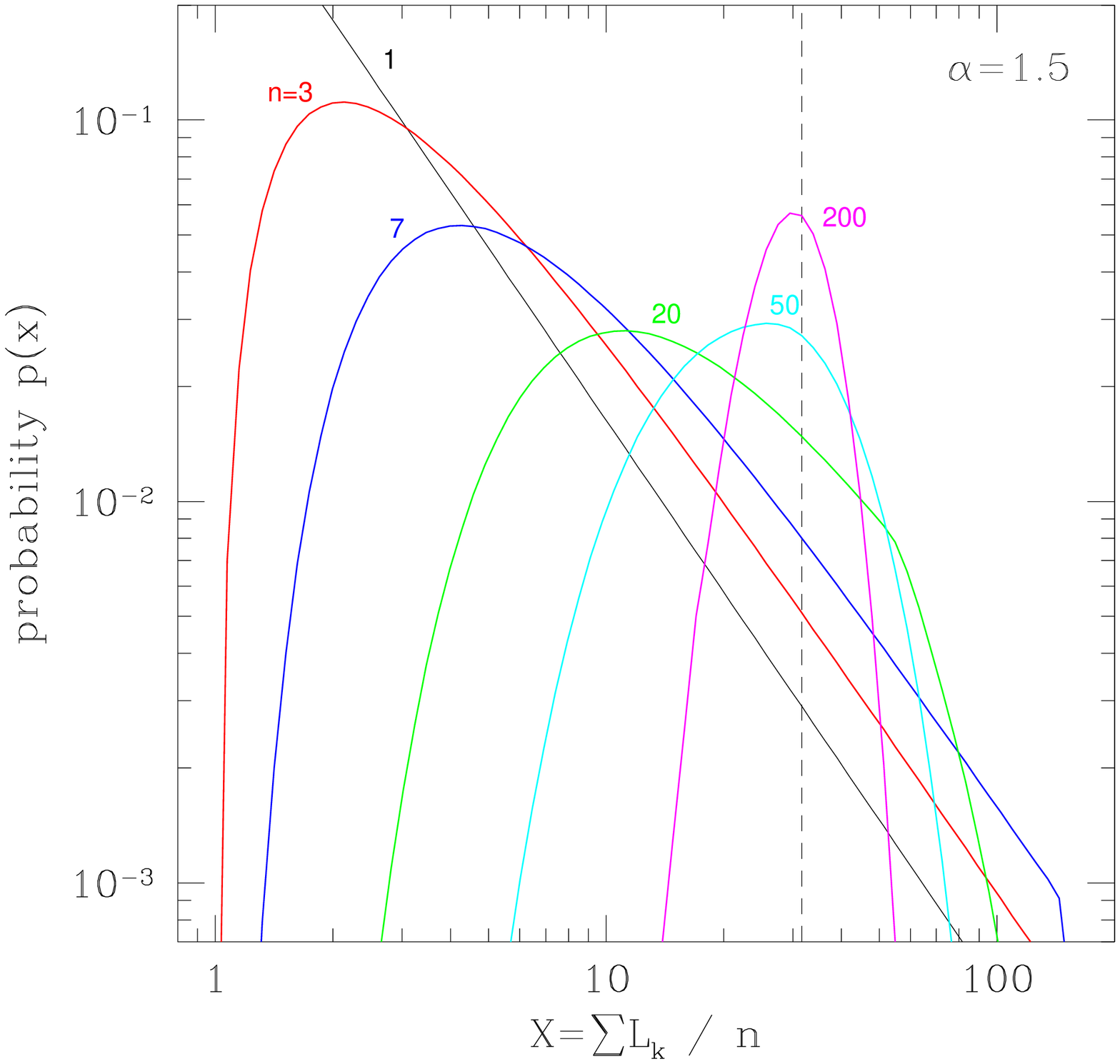}}
}
\hbox{
\resizebox{0.40\hsize}{!}{\includegraphics[bb=18 165 574 705]{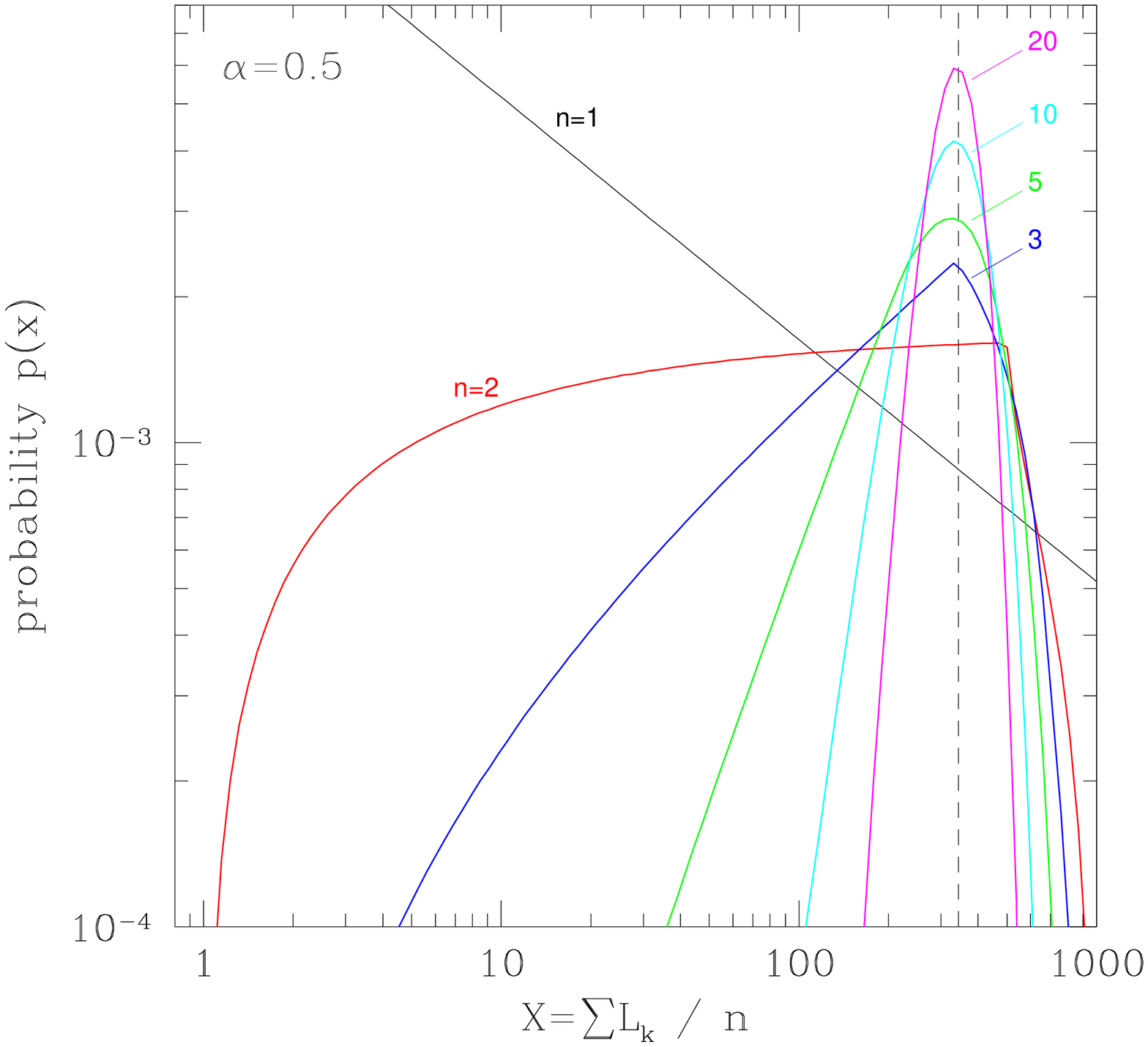}}
\resizebox{0.40\hsize}{!}{\includegraphics[bb=18 165 574 705]{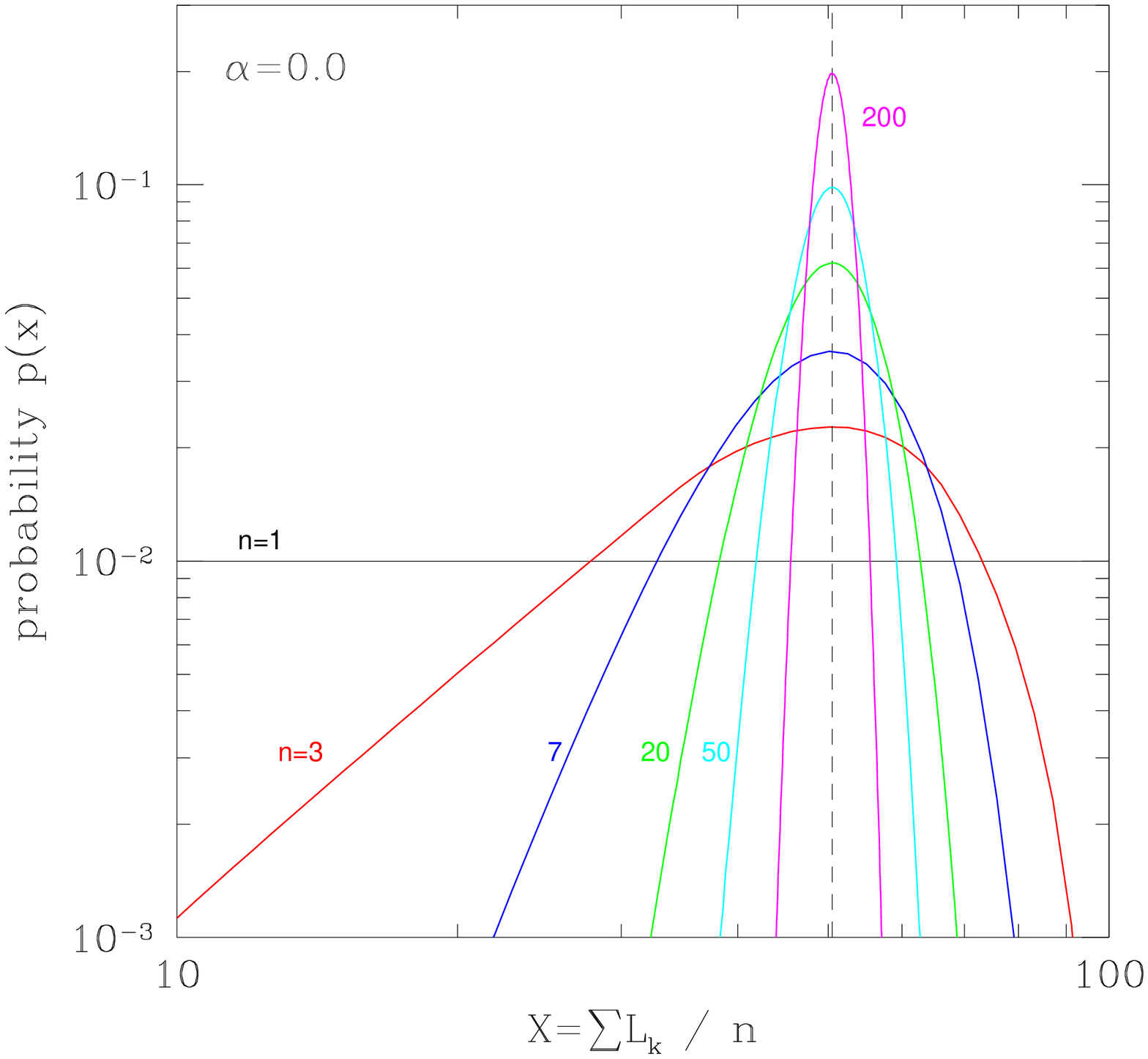}}
}
}
}
\caption{Probability distributions of the average luminosity, 
$\sum_{k=1}^n L_k/n$, of $n$ discrete sources with a power law LF, 
eq.(\ref{eq:lfd_pl}). The lower and upper luminosity cut-offs were
fixed at $L_1=1$ and $L_2=10^3$ for all plots. The value of the slope
$\alpha$ is indicated in each panel. Each curve is marked 
according to the number of sources $n$. The vertical dashed lines
show the expectation mean $\left< L_{\rm tot} \right>/n$. 
}
\label{fig:probdist}
\end{figure*}

\begin{figure*}
\centerline{
\hbox{
\resizebox{0.45\hsize}{!}{\includegraphics{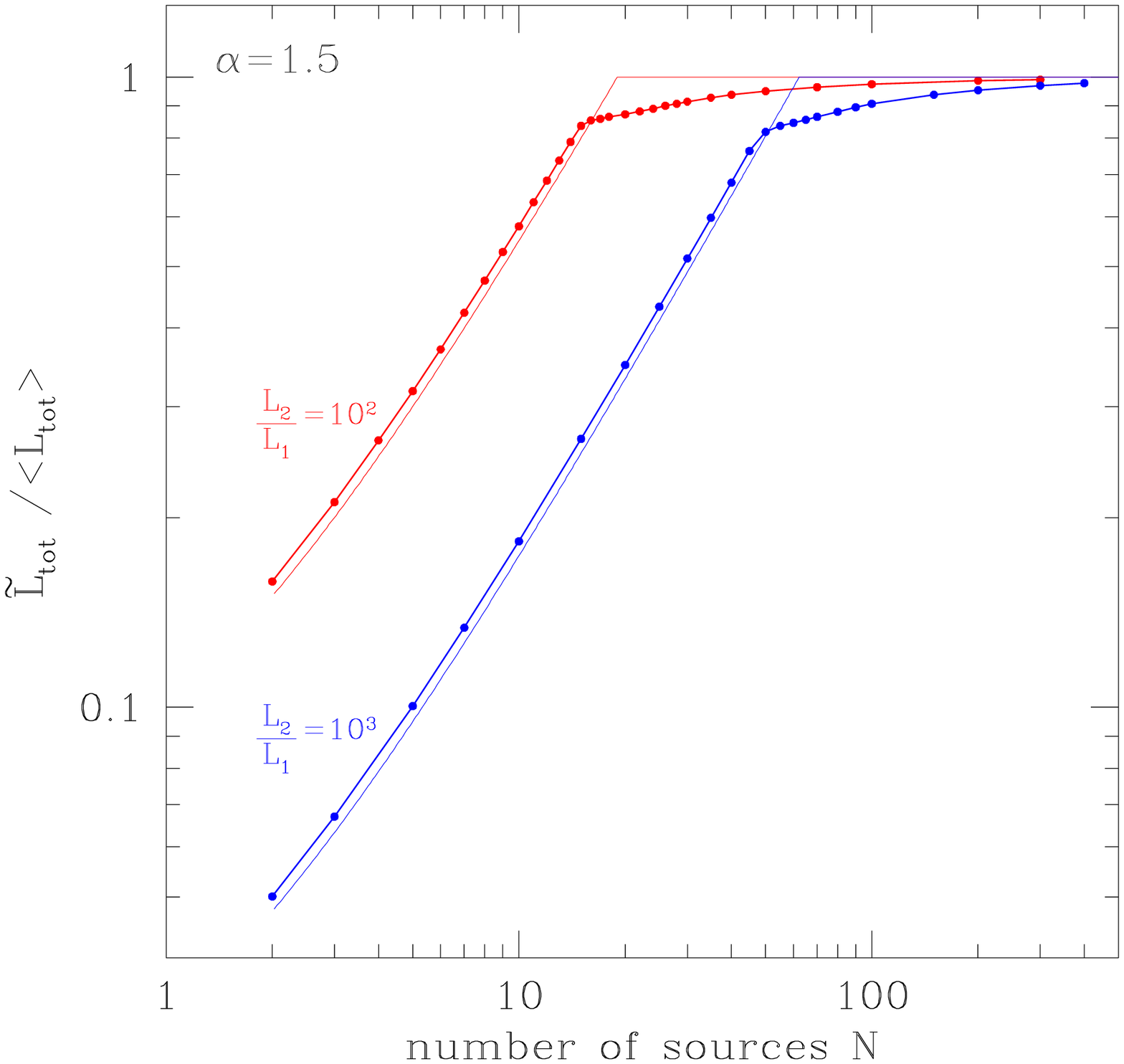}}
\resizebox{0.45\hsize}{!}{\includegraphics{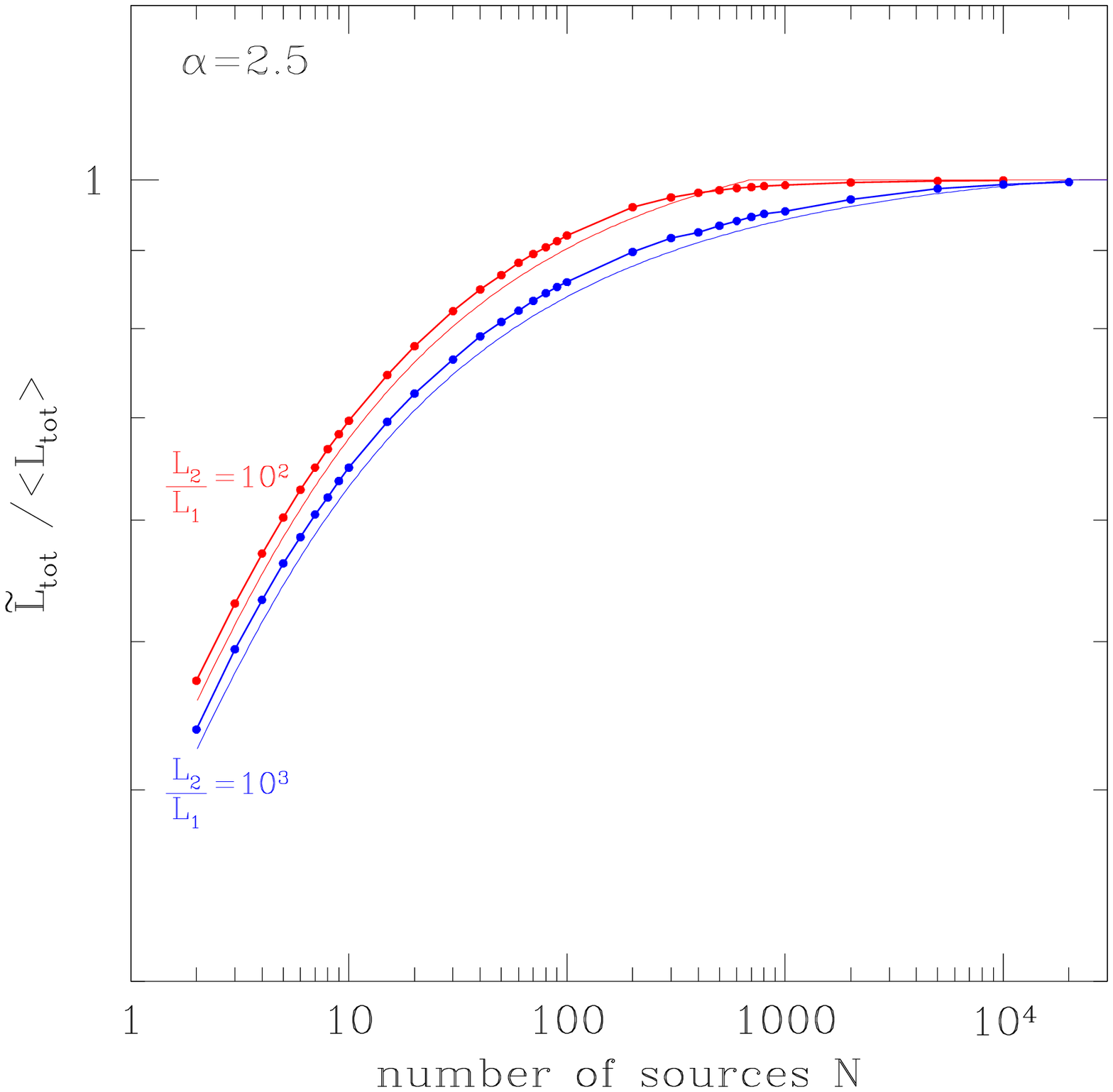}}
}
}
\caption{The ratio of the most probable value of the total luminosity
$\tilde{L}_{\rm tot}$ to the expectation mean
$\left< L_{\rm tot} \right>$ versus number of sources observed in
$L_1-L_2$ luminosity range for different
values of the LF slope $\alpha$ and the ratio
$L_2/L_1$. The results of exact calculation using
eq.(\ref{eq:pn_ltot}) are shown by the solid symbols connected with
thick line. The thin solid line shows the approximate relation calculated
from eqs.(\ref{eq:n_of_xi}) and (\ref{eq:ltot_of_lmax}), as described
in the Appendix \ref{sec:lprob_approx_recipe}.
}
\label{fig:lprobn}
\end{figure*}

\subsubsection{Method II}
\label{sec:method2}

The probability distribution of the total
number of sources follows a Poisson distribution
$P_{\mu}(n)=\mu^n\, e^{-\mu}/n!$ with mean $\mu$ given by 
eq.(\ref{eq:ntot}). 
The probability distribution of the total luminosity is:
\begin{eqnarray}
p\,(L_{\rm tot})=\sum_{n=0}^{n=\infty} P_{\mu}(n)\,
p_n(L_{\rm tot})
\label{eq:pltot_pnltot}
\end{eqnarray}
where $p_n(L_{\rm tot})$ is the probability distribution of the total
luminosity of exactly $n$ sources.
In the majority of practically interesting cases, the total number of
sources is sufficiently large, $\left< n \right> >>1$, and  the
Poisson distribution in eq.(\ref{eq:pltot_pnltot}) can be replaced by
the delta-function  $\delta(n-\left< n \right>)$:  
\begin{eqnarray}
p\,(L_{\rm tot})\approx p_n(L_{\rm tot}), ~~n=\left< n \right>>>1
\label{eq:pltot1}
\end{eqnarray}
where $n=\left< n \right>$ is given by eq.(\ref{eq:ntot})

The probability distribution of the sum of $n$ random variables,
$L_{\rm tot}=\sum_{k=1}^{k=n} L_k$,
can be calculated using the convolution theorem:
\begin{eqnarray}
\widehat{p_n}=\widehat{p_1}^n
\nonumber
\\
\widehat{p_1}(\omega)=\int_0^{\infty} p_1(L)\,e^{i\omega L}\,dL
\label{eq:pn_ltot}
\\
p_n(L_{\rm tot})=\int_{-\infty}^{+\infty} \widehat{p_n}(\omega)\,
e^{-i\omega L_{\rm tot}} \, d\omega
\nonumber
\end{eqnarray}
where $p_1(L)$ is the probability distribution for the luminosity
of one source, which equals the luminosity function with appropriate
normalization: 
\begin{eqnarray}
p_1(L)=\frac{1}{\left< n \right>}\ \frac{dN}{dL}
\label{eq:p1}
\end{eqnarray}


The probability distribution $p_n(L_{\rm tot})$, defined by 
eq.(\ref{eq:pn_ltot}), describes the case when $n$ sources are {\em 
observed}. On the contrary, the distribution $p\,(L_{\rm tot})$,
defined by eq.(\ref{eq:p_ltot}), is parametrized via the normalization of the
luminosity function or, equivalently, via the expectation value
$\left< n \right>$, and describes the case when $\left< n \right>$
sources are {\em expected}. These two distributions are related via
eq.(\ref{eq:pltot_pnltot}). For $n>>1$, which is often the case, they
are nearly identical, and it can be assumed that $n=\left< n \right>$
and both quantities are related to the normalization of the luminosity
function via eq.(\ref{eq:ntot}).  

\begin{figure*}
\centerline{
\hbox{
\resizebox{0.45\hsize}{!}{\includegraphics[bb=18 180 592 700]{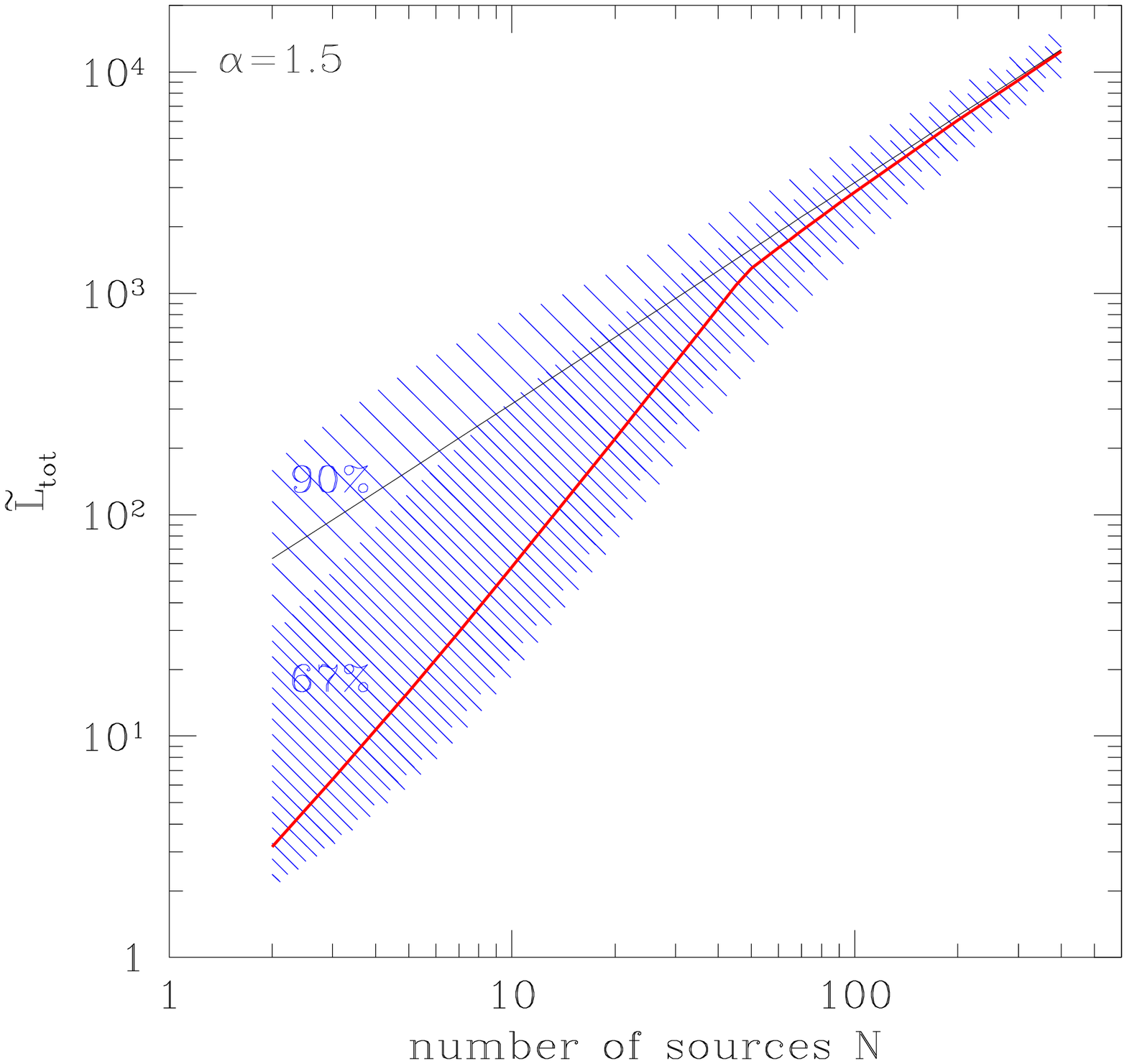}}
\resizebox{0.45\hsize}{!}{\includegraphics[bb=18 180 592 700]{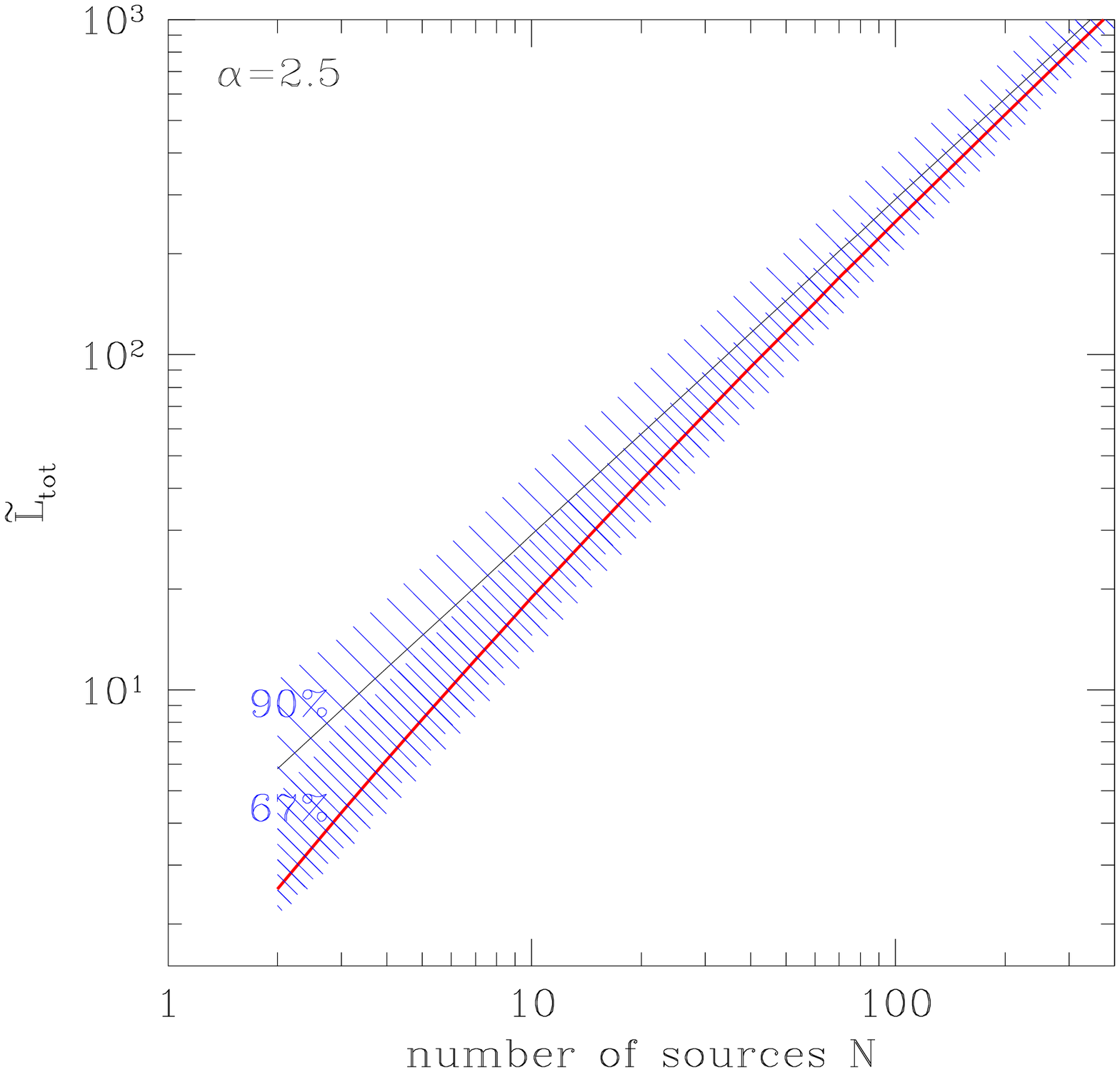}}
}
}
\caption{The most probable value of the total luminosity
$\tilde{L}_{\rm tot}$ and its intrinsic dispersion versus number of
observed sources for different values of the LF slope
$\alpha$.  
The lower and upper  cut-off luminosities were fixed at
$L_1=1$, $L_2=10^3$. The results of exact calculation using
eq.(\ref{eq:pn_ltot}). Behavior of $\tilde{L}_{\rm tot}$ is shown by
the thick solid line. The narrower and broader dashed areas correspond
to its 67\% and 90\% intrinsic dispersion. 
The thin solid line shows linear relation for the expectation mean
$\left<L_{\rm tot} \right>\propto n$. 
}
\label{fig:lprob}
\end{figure*}

\subsection{Illustrative examples}
\label{sec:examples}

To illustrate the behaviour of the total luminosity we assume a power
law luminosity function:
\begin{eqnarray}
\frac{dN}{dL}=\left\{ \begin{array}{ll}
A L^{-\alpha} 	& \mbox{if $L_1\le L \le L_2$}\\
0& \mbox{otherwise}
\end{array} \right.
\label{eq:lfd_pl}
\end{eqnarray}

\subsubsection{Probability distribution $p_n\,(L_{\rm tot})$}
\label{sec:probdist_results}

The probability distributions $p_n(L_{\rm tot})$ for various values of
the slope $\alpha$ and the number of sources $n$ are shown in
Fig.\ref{fig:probdist}. To facilitate the comparison of 
distributions for different values of $n$, the abscissa in these plots
is  the average luminosity  $L_{\rm tot}/n=\sum_{k=1}^{k=n}
L_k/n$. The values of the luminosity cut-offs were fixed at $L_1=1$
and $L_2=10^3$ 

The skewness of the probability distribution $p_n(L_{\rm tot})$ 
leads to a deviation of its mode $\tilde{L}_{\rm tot}$ (the most
probable value of the total luminosity) from the expectation mean
$\left<L_{\rm tot}\right>$ indicated in each panel by the vertical
dashed line. The effect is most pronounced for $\alpha\sim 3/2$, is
unimportant for for shallow luminosity functions with $\alpha<1$ and
gradually diminishes with increasing $\alpha$ at $\alpha>2$.

For illustration, we also show in Fig.\ref{fig:probdist} the
probability distributions for a flat LF,  $\alpha=0$, in which case
$\tilde{L}_{\rm tot}$ 
exactly equals  $\left<L_{\rm tot}\right>$. Naturally, for any value
of $\alpha$ the $p_n(L_{\rm tot})\rightarrow$ Gaussian distribution in
the limit of $n\rightarrow \infty$, in accord with the Central Limit
Theorem. Correspondingly 
$\tilde{L}_{\rm tot}\rightarrow\left<L_{\rm tot}\right>$ in this limit.

\subsubsection{The most probable value of total luminosity}
\label{sec:ltot}

The dependence of the most probable value of the total luminosity
$\tilde{L}_{\rm tot}$ on the total number of sources $n$ for
different values of $\alpha$ and the ratio $L_2/L_1$ is shown
in Fig.\ref{fig:lprobn} and \ref{fig:lprob}. Convenient analytical
approximations for the $\tilde{L}_{\rm tot}-n$ relation are derived in 
Appendix \ref{sec:lprob_approx} and its asymptotical behavior is
considered in Appendix \ref{sec:asympt}.

The $\tilde{L}_{\rm tot}-n$ relation is significantly non-linear
for ``small'' number of sources or, equivalently, for small values of
the LF normalization $A$. For $L_1<<L_2$, $\alpha>1$, the boundary
between the non-linear and linear regimes, expressed in terms of the
normalization $A$, depends only on the LF slope $\alpha$  and its high
luminosity cut-off $L_2$ (see Appendix \ref{sec:asympt}). This is due
to the fact that the behavior of $\tilde{L}_{\rm tot}$ is defined by
the number of sources near the high luminosity cut-off of the
luminosity function, rather than by the total number of sources in the
entire $L_1-L_2$ luminosity range, i.e. the condition for the linear
regime is:
\begin{eqnarray}
N(L\sim L_2)\sim L_2\frac{dN}{dL}\left(L=L_2\right)\ga 1
\end{eqnarray}
The total number of sources in the $L_1-L_2$ luminosity range, on the
contrary, is defined by the low luminosity cut-off $L_1$ (for
$\alpha>1$). For sufficiently large $L_2/L_1$, the  non-linear
regime can occur for an arbitrarily large total number of sources
(cf. the curves corresponding to different values of $L_2/L_1$ in
Fig.\ref{fig:lprobn}). Interestingly, for the LF slope in the range
of $1<\alpha < 2$, where the effect is strongest,
there is a relatively sharp break separating the
non-linear part of the dependence from the linear part.

Although for small $n$ the most probable value of luminosity
$\tilde{L}_{\rm tot}$ can deviate significantly from the expectation
mean $\left< L_{\rm tot} \right>$
(Fig.\ref{fig:probdist},\ref{fig:lprobn}),  the  condition 
$\int p\,(L_{\rm tot})\,L_{\rm tot}\,dL_{\rm tot}=\left< L_{\rm tot}
\right>$ is satisfied for any $n$. Consequently, the average of
$L_{\rm tot}$ over a large number of realizations with the same $n$
always equals $\left< L_{\rm tot} \right>$. 
This equality is achieved due to the existence of outliers, having
values of $L_{\rm tot}$ significantly exceeding both $\tilde{L}_{\rm
tot}$ and $\left<L_{\rm tot}\right>$, in accordance with the skewness
of the probability distribution $p_n(L_{\rm tot})$ for small $n$. This 
naturally leads to enhanced and asymmetric dispersion of the
observed values of $L_{\rm tot}$ in the non-linear regime as
illustrated by the shaded areas in Fig.\ref{fig:lprob}.

\subsection{The luminosity of the brightest source in a sample}
\label{sec:prob_lmax}

The probability distribution for the luminosity of the brightest
source observed in  a sample of $n$ sources equals:   
\begin{eqnarray}
p\,(L_{\rm max})= \left[ p_1(L<L_{\rm max}) \right]
^{n-1} p_1(L_{\rm max})\ n
\label{eq:p_lmax}
\end{eqnarray}
where $p_1(L)$ is the probability distribution for the luminosity of
one source (e.g. eq.(\ref{eq:p1})) and
$p_1(L<L_{\rm max})$ denotes the cumulative probability 
$p_1(L<L_{\rm max})=\int_0^{L_{\rm max}} p_1(L)\,dL$. 

\begin{figure}
\resizebox{\hsize}{!}{\includegraphics{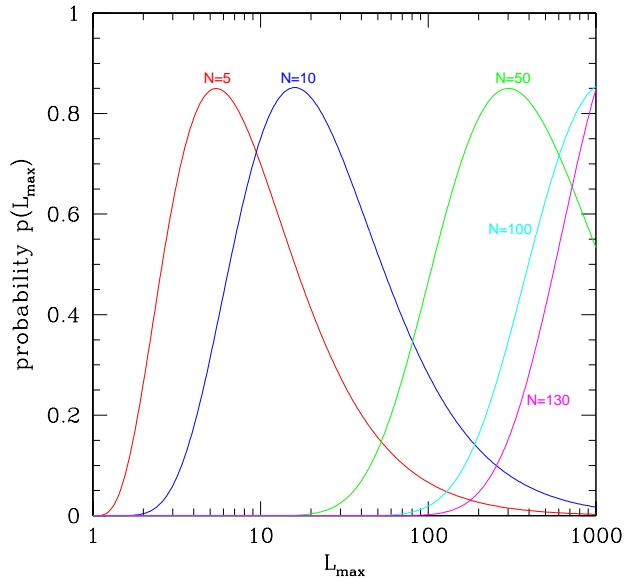}}
\caption{Probability distributions $p\,(L_{\rm max})$ 
of the luminosity of the brightest source in a
sample of $n$ sources with a power law LF with the slope $\alpha=1.5$
and $L_1=1$, $L_2=10^3$. Each curve is marked according to the number
of sources $n$.   
}
\label{fig:prob_lmax}
\end{figure}

The $p\,(L_{\rm max})$ distribution for $\alpha=1.5$ 
is shown in Fig.\ref{fig:prob_lmax} and illustrates the intuitively
obvious fact that if the number of sources is sufficiently small the
brightest sources most likely will not reach the highest possible
value of $L_2$. An analytical formula for the most probable value of
the luminosity of the brightest source is given by eq.(\ref{eq:lmax}).

\begin{figure*}
\centerline{
\hbox{
\resizebox{0.45\hsize}{!}{\includegraphics{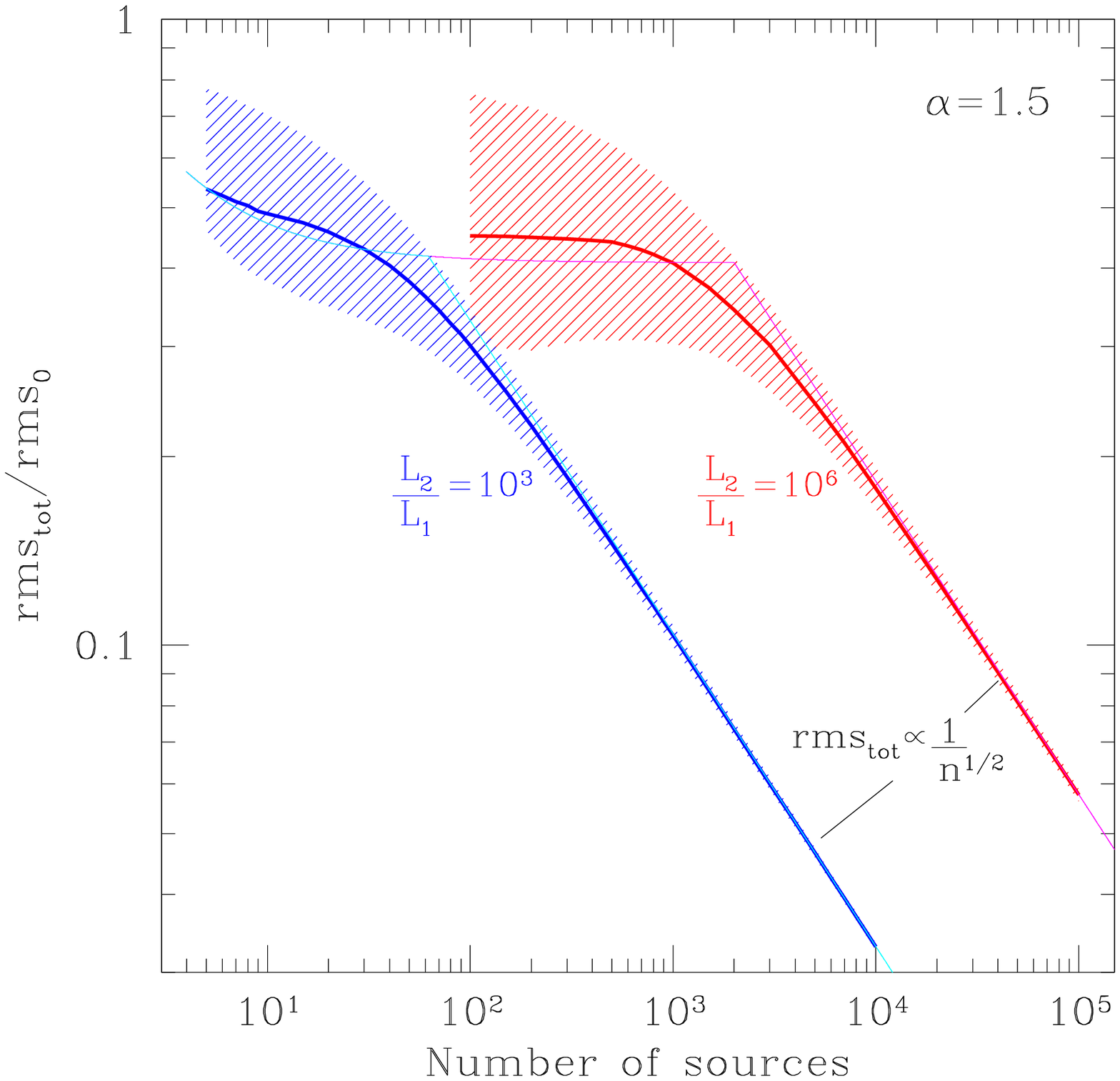}}
\resizebox{0.45\hsize}{!}{\includegraphics{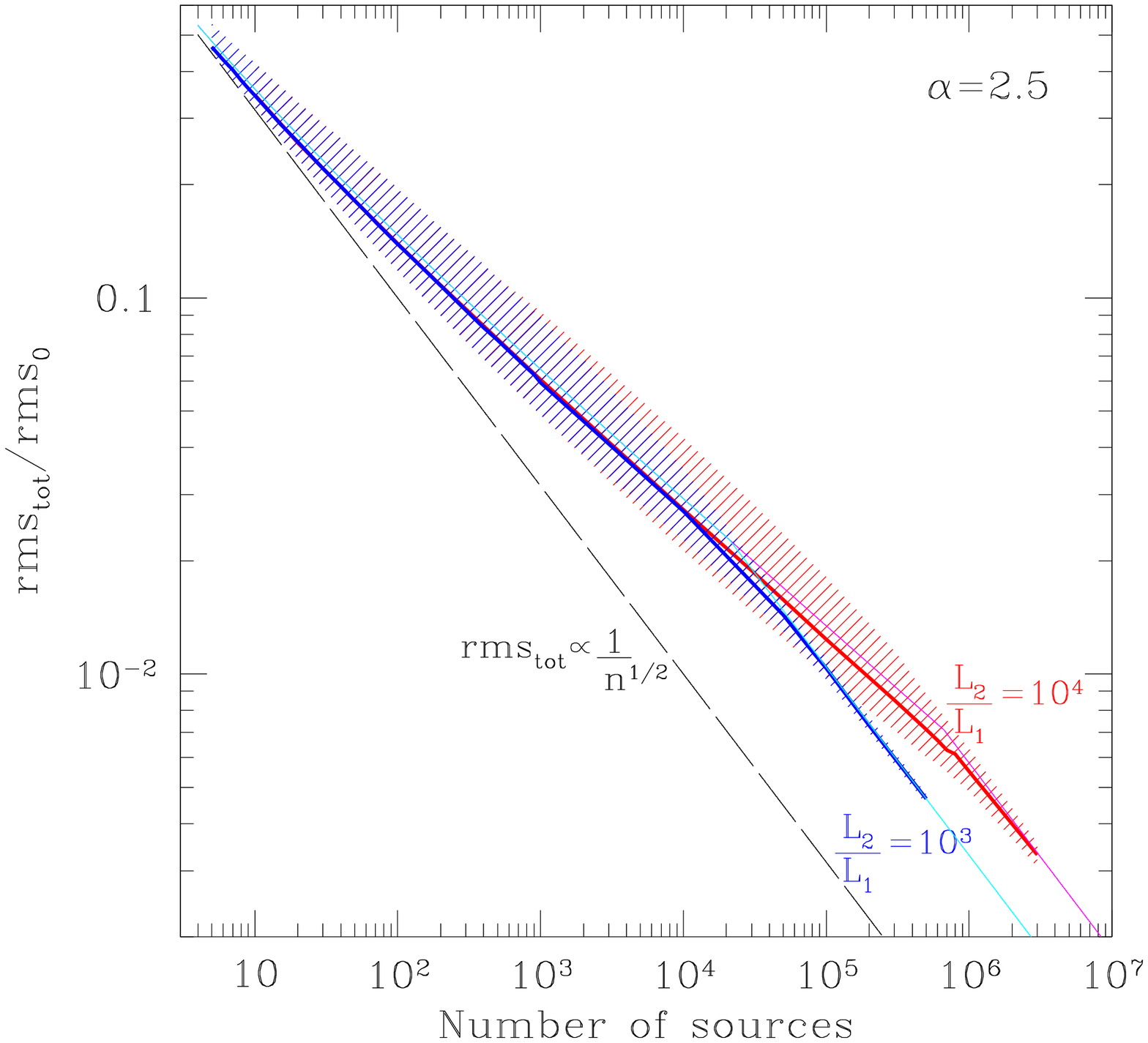}}
}
}
\caption{Variability of the total emission. The ratio of
the fractional $rms$ of the combined emission to that 
of one source, $rms_0$, as a function of the number of sources for
different values  
of the LF slope $\alpha$ (indicated in the
upper-right corner of each plot) and different values of the ratio of
$L_2/L_1$. The thick solid lines show the square root of the most
probable value of the $rms_{\rm\,tot}^2/rms_0^2$, 
the shaded areas show the 67\% intrinsic
dispersion, both obtained from the Monte-Carlo simulations. The thin
solid lines were computed using the approximation given by 
eq.(\ref{eq:rmstot_of_n_small_n}). The dashed line shows
$rms_{\rm\, tot}/rms_0\propto1/\sqrt{n}$ dependence.
}
\label{fig:var}
\end{figure*}

\section{Variability of the total emission}
\label{sec:variability}

For a population of $n$ sources with luminosities $L_k$
and fractional $rms$ of aperiodic variability $rms_k$, the fractional
rms of the total emission  equals:
\begin{eqnarray}
rms_{\,\rm tot}^2=\frac{\sum_{k=1}^n L_k^2\ rms_k^2}
{\left(\sum_{k=1}^n L_k\right)^2}
\label{eq:rmstot}
\end{eqnarray}
assuming that variations of the source fluxes are uncorrelated with
each other. In the following we also assume for simplicity that all
sources have the same value of fractional rms, i.e.
$rms_k=rms_0$.

In the limit of $n\rightarrow\infty$, corresponding to the
linear regime in the $\tilde{L}_{\rm tot}-n$ relation, the sums in the 
eq.(\ref{eq:rmstot}) can be replaced by the respective integrals of
the LF:
\begin{eqnarray}
\frac{rms_{\,\rm tot}^2}{rms_0^2}
=\frac{\int L^2\,dN/dL\,dL}
{\left(\int L \,dN/dL\,dL \right)^2}
\propto \frac{1}{n} \propto \frac{1}{A} \propto \frac{1}{L_{\rm tot}}
\label{eq:rmstot_large_n}
\end{eqnarray}
In the linear regime the fractional $rms$ of the collective
emission obeys $\propto 1/\sqrt{n}$ averaging law, as
expected for uncorrelated variations of individual sources.

In the non-linear regime, however, for a sufficiently flat luminosity
function, the total luminosity is defined by a few brightest sources.
To first approximation the number of such sources effectively
contributing to the sums in  eq.(\ref{eq:rmstot}) does not depend on
the LF normalization.
Consequently, the fractional $rms$ of the total emission is constant,
independently on the total number of sources or of their total
luminosity. This contradicts the intuitive expectation that the
fractional $rms$ decreases with the number of sources as  $rms\propto
1/\sqrt{n}$.

To illustrate this behavior we performed a series of the Monte-Carlo
simulations for a power law LF. For each set of parameters and a given 
value of the number of sources, in each run $n$ sources were 
placed between $L_1$ and $L_2$ with a power law luminosity
distribution, eq.(\ref{eq:lfd_pl}). For each run, the ratio 
$rms_{\,\rm tot}^2/rms_0^2$ was computed following
eq.(\ref{eq:rmstot}). From results of many runs, the probability
distribution for $rms_{\,\rm tot}^2/rms_0^2$ was obtained. The maximum
of this distribution defines the most probable value of 
$rms_{\,\rm tot}^2$ and its width characterizes the intrinsic
dispersion of this quantity. 
The examples are shown in Fig.\ref{fig:var} for two values of the LF
slope, $\alpha=1.5,\, 2.5$. For $\alpha<1$
and $\alpha>3$ there is no non-linear regime and the 
fractional $rms$ of the total emission obeys $\propto 1/\sqrt{n}$ law
for any $n$. 
In plotting Fig.\ref{fig:var}, we converted $rms_{\,\rm tot}^2$ to 
$rms_{\,\rm tot}$, so that the thick solid curves shows the behavior
of the square root of the most probable value of the 
$rms_{\,\rm tot}^2/rms_0^2$ ratio.

Approximate formulae for $rms_{\,\rm tot}^2$ are derived in Appendix
\ref{sec:variability_ap}. 
\begin{figure*}
\centerline{\hbox{
\resizebox{0.45\hsize}{!}{\includegraphics{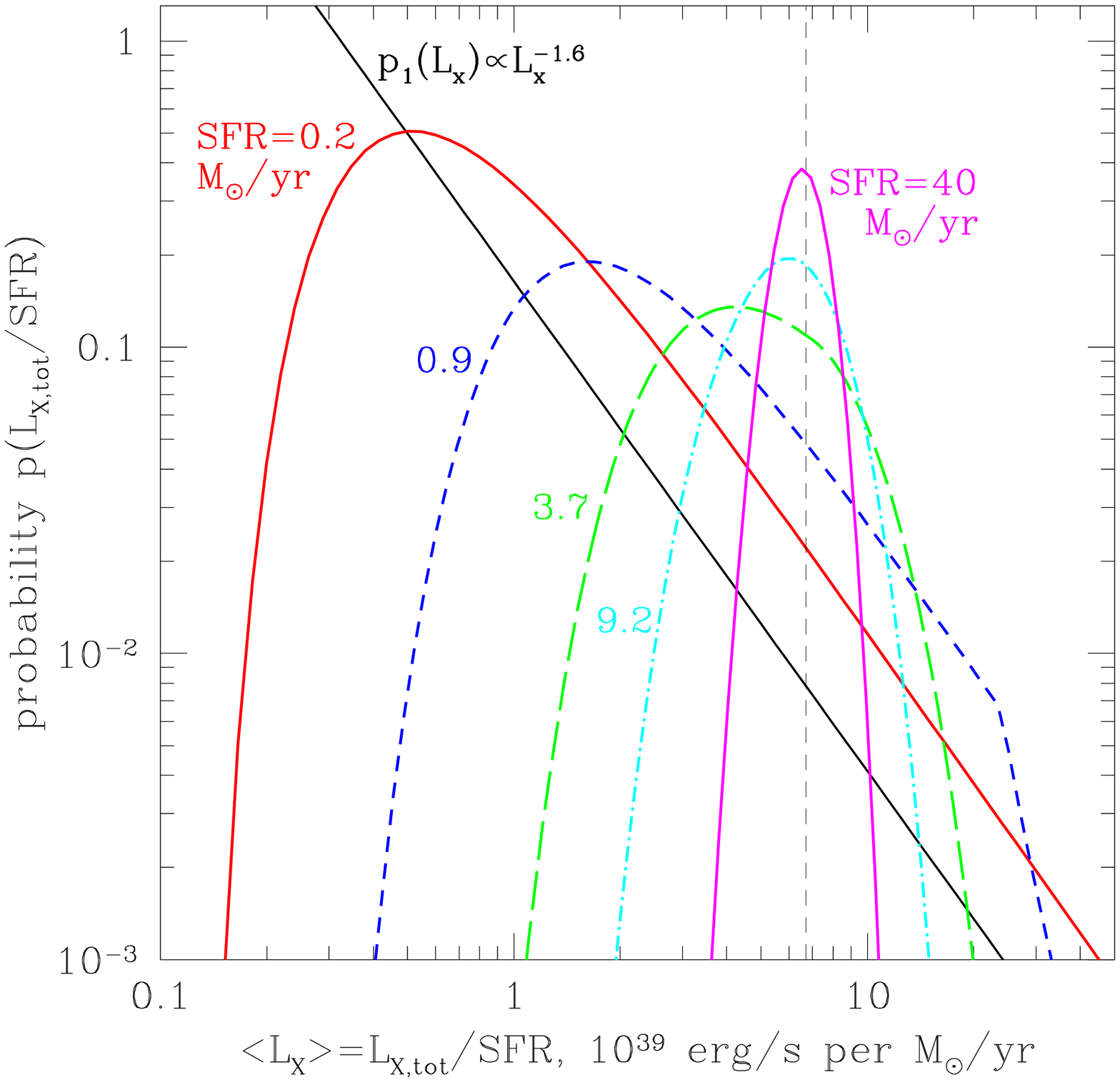}}
\resizebox{0.45\hsize}{!}{\includegraphics{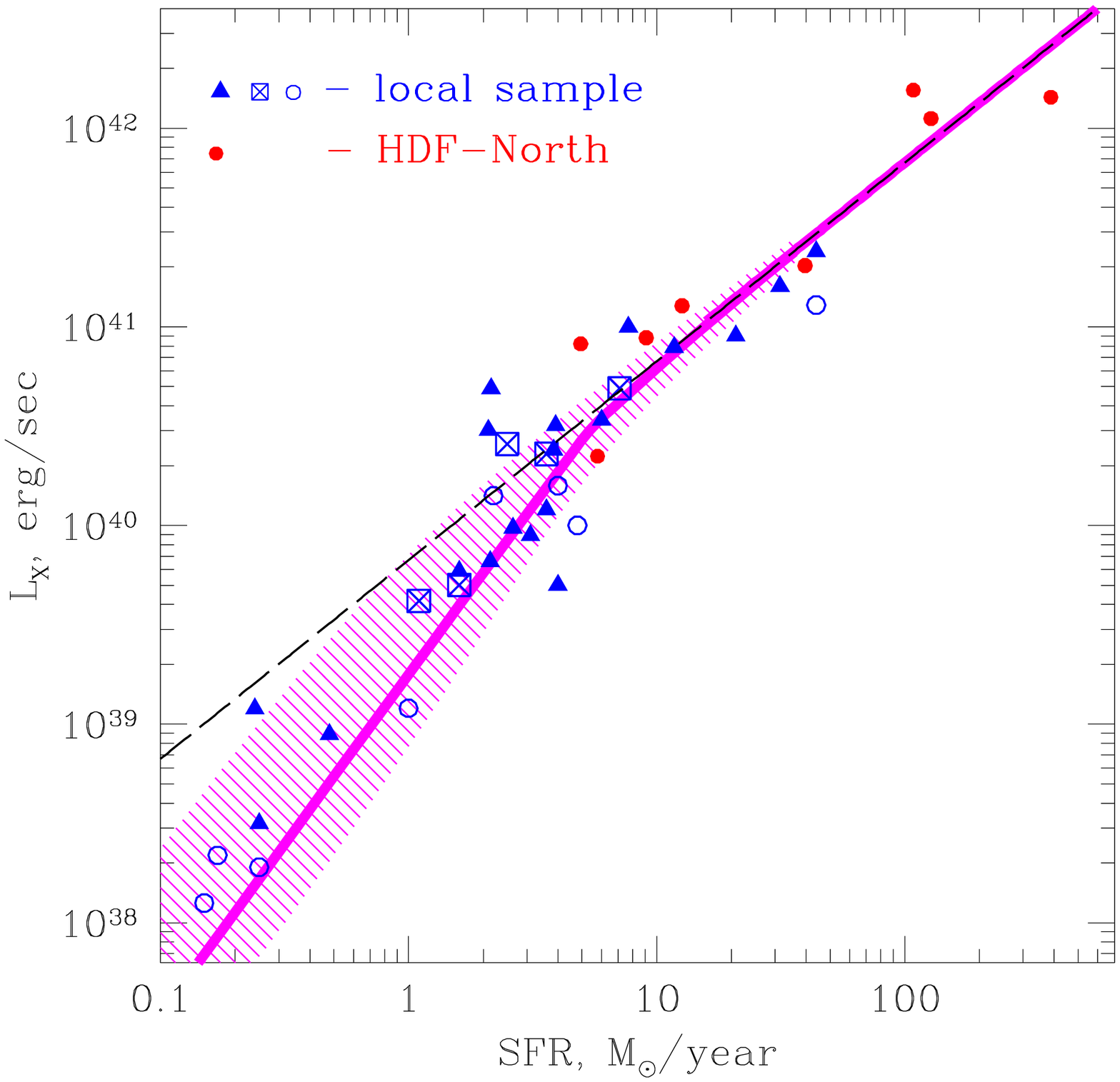}}
}}
\caption{
{\em Left:} The probability distributions $p(L_{\rm tot}/{\rm SFR})$
for different values of SFR. The vertical dashed line shows the
expectation mean, defined by eq.(\ref{eq:ltot_mean}).
{\em Right:} The $L_X-$SFR relation. The open circles are nearby 
galaxies observed by Chandra, the filled triangles are spatially
unresolved nearby galaxies observed by ASCA and BeppoSAX, for which
only total luminosity is available, the filled circles are distant
star forming galaxies from  the HDF-N. The thick
grey line is relation between the SFR and the most
probable value of the total luminosity, predicted from the
``universal'' HMXB XLF, the shaded area shows its 67\%
intrinsic spread, the dashed line is the
expectation mean, defined by eq.(\ref{eq:ltot_mean}). 
The five nearby galaxies, used by \citet{grimm} to derive the
``universal'' HMXB XLF, 
are marked as crossed boxes. 
} 
\label{fig:lx-sfr}
\end{figure*}

\section{Astrophysical applications}
\label{sec:applications}

\subsection{Determining LF parameters from total emission of
unresolved sources} 
\label{sec:lf_parameters}

The shape of the $\tilde{L}_{\rm tot}-A$ relation ($A$ is the
normalization of the luminosity function) is defined by the
parameters of the luminosity function. For $1\la\alpha\la 2$, it has
two distinct power law regimes, separated by a
break (Fig.\ref{fig:lprobn} and \ref{fig:lprob}):
\begin{eqnarray}
\renewcommand{\arraystretch}{1.3}
\tilde{L}_{\rm tot} \propto \left\{ \begin{array}{ll}
A^{\frac{1}{\alpha-1}}& A<A_{\rm break}\\
A	& A>A_{\rm break}
\end{array}
\right.
\label{eq:ltot_of_a}
\end{eqnarray}
The position of the break between the non-linear and linear regime is
defined by the high luminosity cut-off of the LF and its slope as
described by eq.(\ref{eq:abreak}). This opens a possibility to
determine the LF parameters without actually resolving the individual
sources, but studying large samples of objects (e.g. galaxies) and the
relation between their total emission and the normalization $A$. The
value of $A$ in many cases can be determined independently from
observations at other wavelengths.
For example, the normalizations of the luminosity functions of high
and low mass X-ray binaries are proportional to the star formation rate
\citep{grimm} and stellar mass \citep{lmxb} of the host galaxy
respectively. Both quantities can be determined from the conventional
indicators, such as radio or near-infrared luminosities.

Of course with the sub-arcsec angular resolution of Chandra the
luminosity distribution of point sources in nearby galaxies
can be studied directly. However, for more distant galaxies, 
D$\ga 30-50$ Mpc, even Chandra resolution becomes insufficient and
only the total luminosity of the galaxy can be measured. Provided that
the contaminating contribution of the emission of a central AGN and/or
of hot X-ray emitting gas can be constrained or separated, one can
study the relation between the total luminosity $L_X$ of a
(unresolved) galaxy and its star formation rate or stellar mass. The
$L_X-$SFR or $L_X-M_*$ ``growth curves'' constructed for large samples
of galaxies and spanning a broad range of SFR and $M_*$ can be used to 
constrain the XLF parameters of X-ray binaries in distant galaxies
including those located at intermediate and high redshifts. 
With this, one can study the influence of a number of factors such as
effects of binary evolution, metallicity, regimes of star formation,
etc. on the luminosity distribution of X-ray binaries.

\subsection{High mass X-ray binaries in star forming galaxies}
\label{sec:hmxb}

\subsubsection{$L_X-$SFR relation for star forming galaxies}

The slope of the ``universal'' luminosity function of HMXBs,
$\alpha=1.6$, is in the range where the non-linear behavior of 
the $\tilde{L}_{\rm tot}-A$ relation is most pronounced. In the left panel
of Fig.\ref{fig:lx-sfr} we plot the probability distribution of the total
luminosity $p\,(L_{\rm tot})$, computed for different values of the star
formation rate. The distribution is strongly asymmetric in the
non-linear low-SFR regime which, for the parameters of the
``universal'' HMXB XLF, corresponds to the formation rate of massive
stars below SFR$\la 4-5$ M$_{\odot}$/yr. Note that a small value of
SFR does not necessarily imply a small total number of sources, which
is defined by the (unknown) low luminosity cut-off of the HMXB
XLF. For example, for SFR=0.2 M$_{\odot}$/yr, when the non-linear
effect is very pronounced, the total number of sources might be as
large as $\sim 300$ ($\sim 1200$) for the low luminosity cut-off of
$10^{34}$ ($10^{33}$) erg/s. These low values of the star formation
rate correspond to the familiar examples of the Milky Way galaxy and
the Magellanic Clouds. On the opposite end among the relatively
nearby and well known galaxies are the Antennae interacting galaxies
which, with the star formation rate of  SFR$\sim 7$ M$_{\odot}$/yr,
have a nearly symmetric $p\,(L_{\rm tot})$ distribution, sufficiently
close to the normal distribution.

The predicted $L_X-$SFR relation is shown in the right panel
in Fig.\ref{fig:lx-sfr}, along with the measured values of X-ray
luminosities and the star formation rates for a number of nearby
galaxies and galaxies observed with Chandra at intermediate redshifts,
$z\sim 0.2-1.3$, in the Hubble Deep Field North. 
The data shown in
Fig.\ref{fig:lx-sfr} are from \citet{grimm}, complemented with the
local galaxies data from \citet*{ranalli}. In plotting the latter we
removed the duplications and the galaxies likely to be contaminated by
the contribution of low mass X-ray binaries, unrelated to the current
star formation activity, as discussed by \citet{lx-sfr}. The
luminosities and star formation rates for the  Hubble Deep Field North
galaxies \citep{brandt01} were computed for the following cosmological
parameters: $H_0=70$ km/s/Mpc, $\Omega_m=0.3$, $\Lambda=0.7$, 
as described in \citet{lx-sfr}.

Fig.\ref{fig:lx-sfr} further illustrates the difference between the
mode and the expectation mean of the $p\,(L_{\rm tot})$ probability
distribution. The thick solid line in  the right panel shows the
SFR--dependence of the  mode of  $p\,(L_{\rm tot})$ and predicts the
{\em most probable} value of the X-ray  luminosity of a randomly
chosen galaxy. If observations of many (different) galaxies with
similar values of SFR are performed, the obtained values of $L_{\rm
tot}$ will obey the probability distribution depicted in the left
panel. The average of the measured values of $L_{\rm tot}$ will be
equal to the expectation mean given by eq.(\ref{eq:ltot_mean}) and
shown by the dashed straight lines in the left and right panels. 
Due to the properties of $p\,(L_{\rm tot})$ these two quantities are
not identical in the low-SFR limit when the total luminosity is
defined by the small number of the most luminous sources. Only in the
large SFR limit when there are sufficiently many sources with
luminosities near the cut-off of the luminosity function,
$\log(L_X)\sim 40$, does the $L_X-$SFR relation behave in the
intuitively expected way.

Owing to the skewness of the probability distribution $p\,(L_{\rm tot})$
large and asymmetric dispersion among the measured values of $L_{\rm
tot}$ is expected in the non-linear low-SFR regime. This asymmetry is
already seen  in Fig.\ref{fig:lx-sfr} -- at low SFR values there are
more points above the thick solid curve than below. Moreover, the
galaxies lying significantly above the solid and dashed curves in
Fig.\ref{fig:lx-sfr} should be expected at low SFR and will
inevitably appear as the plot is  populated with more objects.  
Such behavior differs from a typical astrophysical situation and
should not be ignored when analyzing and fitting the $L_X$--SFR
relation in the low SFR regime. In particular, due to non-Gaussianity
of the $p\,(L_{\rm tot})$ distribution the standard data analysis
techniques -- least square and $\chi^2$ fitting -- become inadequate.

\begin{figure}
\resizebox{\hsize}{!}{\includegraphics{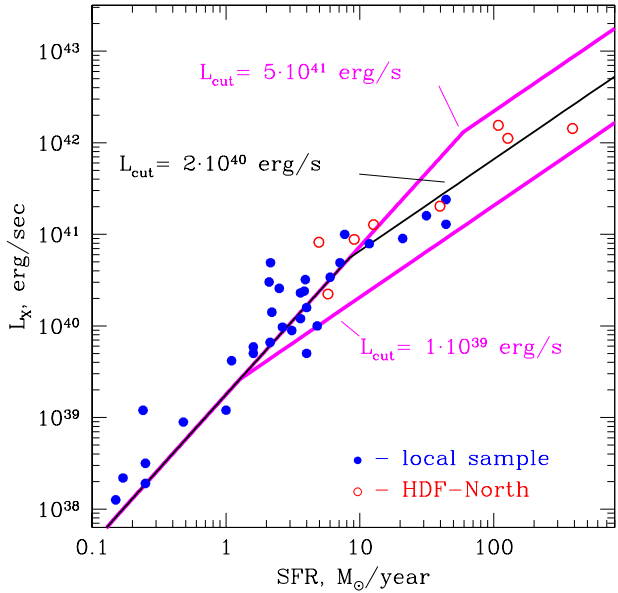}}
\caption{Dependence of the $L_X-$SFR relation on the XLF cut-off
luminosity $L_{\rm cut}$. The three curves,
corresponding to different values of $L_{\rm cut}$ coincide in the
non-linear low SFR regime but differ in the position of the break
between linear and non-linear regimes.
The data are the same as in Fig.\ref{fig:lx-sfr}.
}
\label{fig:lx-sfr-break}
\end{figure}

\begin{figure}
\centerline{
\resizebox{\hsize}{!}{\includegraphics{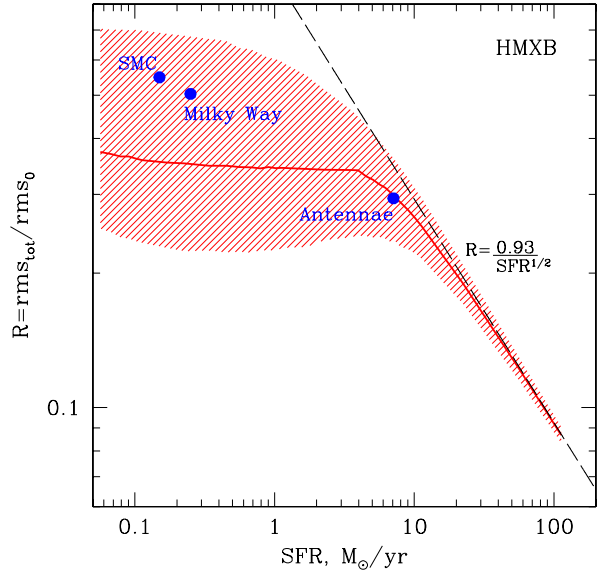}}
}
\caption{Variability of the total emission of HMXBs
in star forming galaxies. Dependence of the ratio 
$rms_{\rm\,tot}/rms_0$ on the star formation rate.
The thick solid line shows the most probable value of the
$rms_{\rm\,tot}/rms_0$, the shaded area shows its 67\% intrinsic
dispersion, both obtained from the Monte-Carlo simulations for a power
law LF with the slope of $\alpha=1.6$ and a cut-off at
$L_2=2\cdot 10^{40}$ erg/s. The dashed line shows asymptotical
behavior at large SFR. The filled circles correspond to HMXB sources
in the Milky Way, SMC and the Antennae galaxies, computed from
eq.(\ref{eq:rmstot}), using the observed luminosities of X-ray sources
in these galaxies. 
}
\label{fig:rms_hmxb}
\end{figure}

\begin{figure*}
\centerline{
\resizebox{0.5\hsize}{!}{\includegraphics{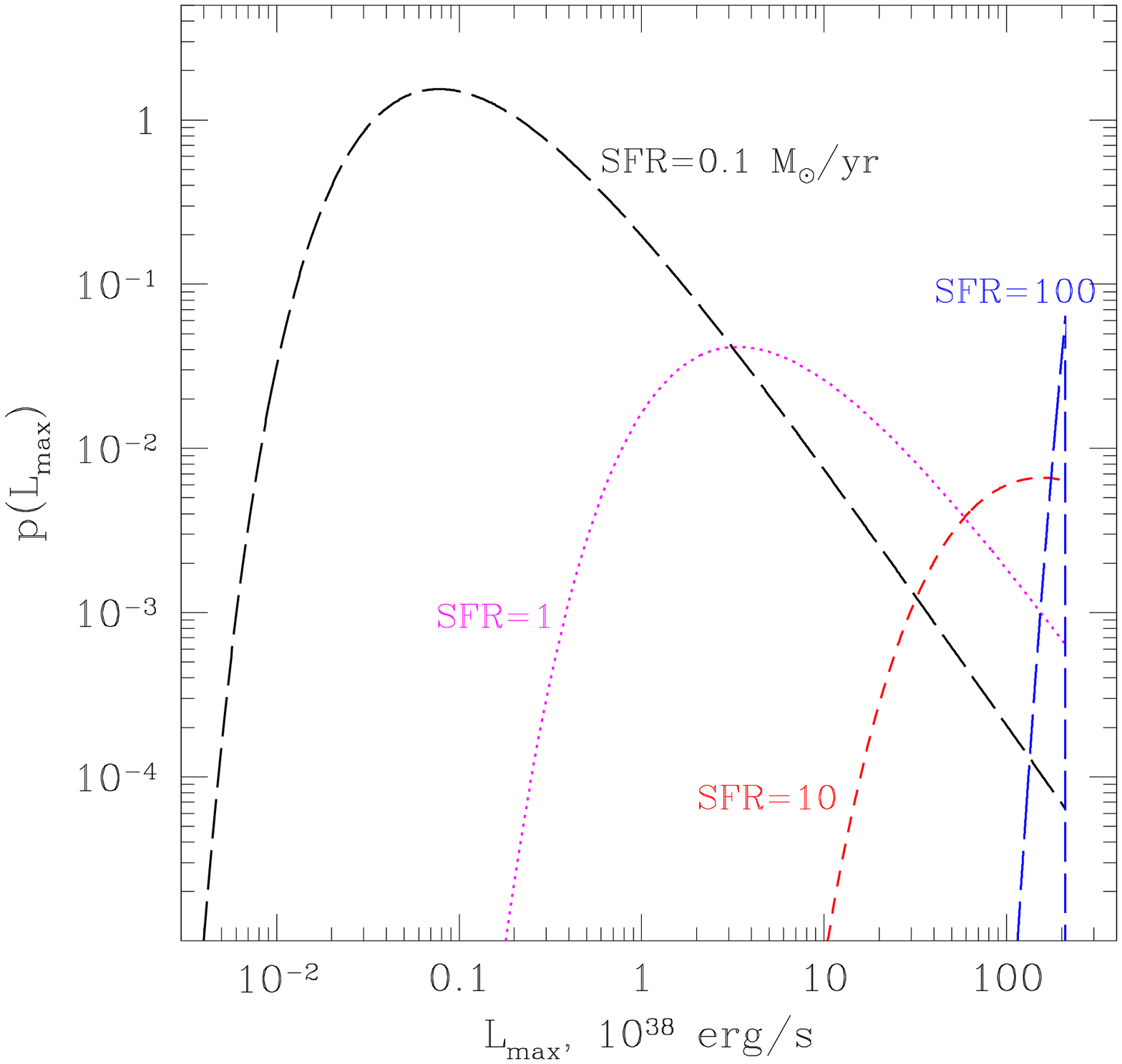}}
\resizebox{0.5\hsize}{!}{\includegraphics{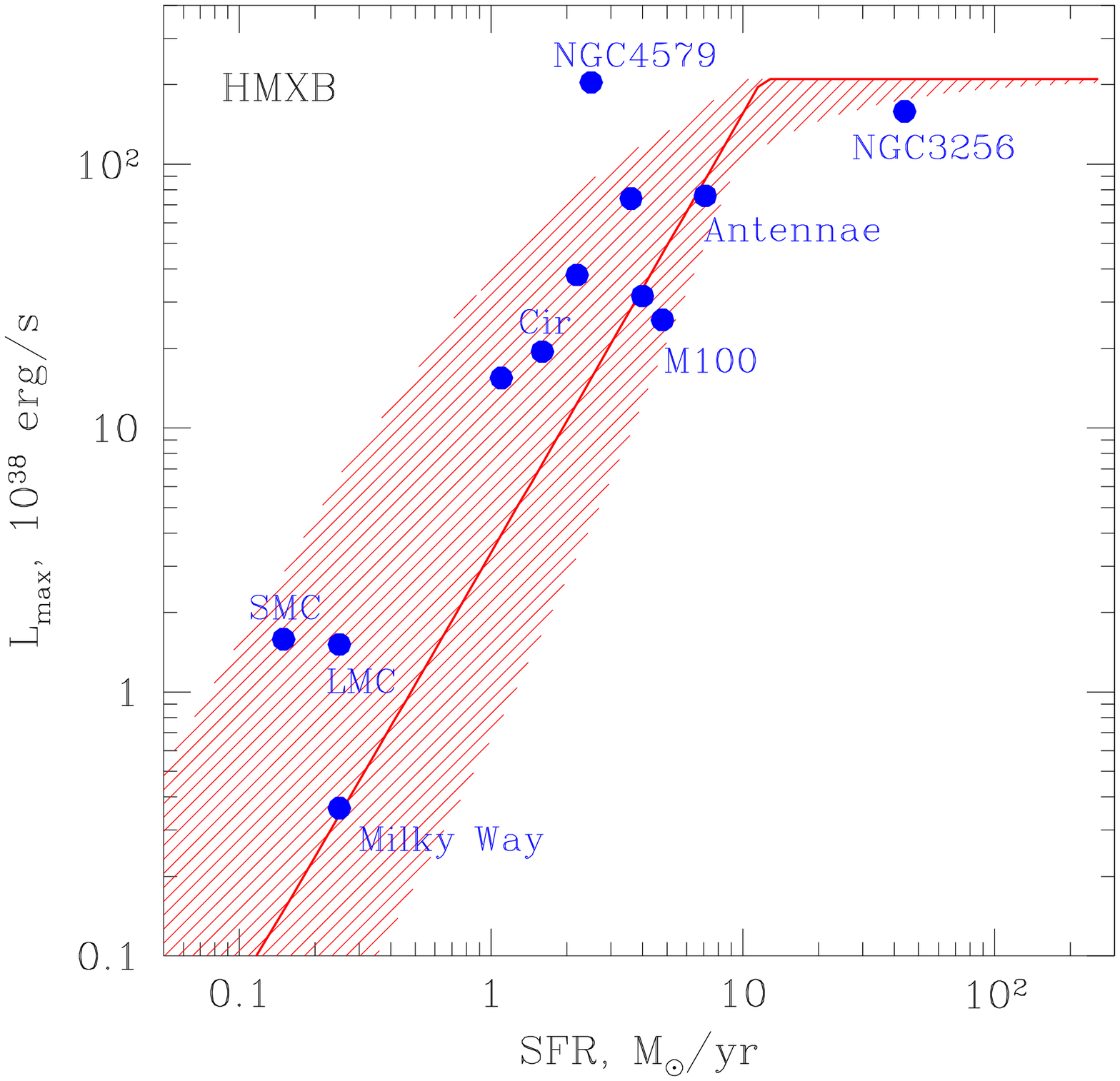}}
}
\caption{{\em Left:} The probability distribution of the luminosity
of the brightest HMXB source for different values of the star
formation rate, computed from eq.(\ref{eq:p_lmax}) using the
parameters of the ``universal'' luminosity function of HMXBs.  
{\em Right:} Expected luminosity of the brightest
HMXB vs. star formation rate. The thick solid line shows the
most probable value of the $L_{\rm max}$ (eq.(\ref{eq:lmax})), 
the shaded area shows its
67\% intrinsic dispersion,  obtained from the probability
distribution given by eq.(\ref{eq:p_lmax}).
The filled circles show maximum observed luminosity of HMXBs in
the Milky Way and several nearby star forming galaxies. 
}
\label{fig:lmax_hmxb}
\end{figure*}

\subsubsection{High luminosity cut-off in the HMXB XLF} 
\label{sec:cutoff_hmxb}

The position of the break between the non-linear and linear parts of
the $L_X$--SFR relation depends on the LF slope and its cut-off
luminosity (Fig.\ref{fig:lx-sfr-break}, eq.\ref{eq:abreak}): 
${\rm SFR}_{\rm break}\propto L_{\rm cut}^{\alpha-1}$. 
This allows to constrain the parameters of the luminosity distribution 
of compact sources using the data of spatially unresolved galaxies as
discussed in section \ref{sec:lf_parameters}. 

Agreement of the predicted $L_X-$SFR relation with the data both
in high- and low-SFR regimes confirms the universality of the HMXB
luminosity function, derived by \citet{grimm} from significantly fewer
galaxies (shown as crossed boxes) than plotted in
Figs.\ref{fig:lx-sfr}, \ref{fig:lx-sfr-break}. It provides an
independent confirmation of the existence of a cut-off in the HMXB
XLF at $\log(L_{\rm cut})\sim 40.5$, including HMXBs in spatially
unresolved high redshift galaxies from the Hubble Deep Field North.
This implies, in particular, that ultra-luminous X-ray sources
(ULX) at redshifts of $z\sim 0.2-1.3$ were not significantly more
luminous than those observed in the nearby galaxies.

\subsubsection{Aperiodic variability}

X-ray flux from X-ray binaries is known to be variable in a broad
range of time scales, from $\sim$msec to $\sim$ years.
In addition to a number of coherent phenomena and 
quasi-periodic oscillations, significant continuum  
aperiodic variability is often observed. The fractional rms of
aperiodic variations depends on the nature of the binary system and
the spectral state of the X-ray source and is usually in the range
from  a fraction of a per cent to $\sim 20-30$ per
cent. Correspondingly, the combined emission of X-ray binaries in a
galaxy should be also variable in a similarly broad range of time scales.  
It has been suggested by \citet*{variability} that due to a large
difference in the characteristic time scales of the accretion flow
onto a stellar mass object and onto a supermassive black hole
variability of the X-ray emission from a galaxy can be used to
distinguish between the combined emission of a population of X-ray
binaries and that of an accreting supermassive black hole in the
centre of a galaxy (AGN).

From results of section \ref{sec:variability} and Appendix
\ref{sec:variability_ap} one should expect that in the non-linear
low-SFR regime the fractional $rms$ of the total X-ray emission
of a star forming galaxy is independent of SFR. 
This prediction is confirmed by the results of the Monte-Carlo
simulations performed as described in section \ref{sec:variability} 
and shown in Fig.\ref{fig:rms_hmxb}. For moderate star formation rates, 
SFR$\la 5-10~M_{\sun}$/yr, we predict a rather large aperiodic
variability of the total emission of HMXBs at the level of $\sim
1/3-1/2$ of the fractional $rms$ of individual X-ray binaries. 
At larger values of SFR, corresponding to the linear regime in the
$L_X-$SFR relation, it decreases as $rms\propto 1/\sqrt{\rm SFR}$, in  
accord with eq.(\ref{eq:rmstot_of_n_small_n}). Also shown in
Fig.\ref{fig:rms_hmxb} are values of the fractional $rms$ reduction,
computed directly from eq.(\ref{eq:rmstot}) using the luminosities of 
the observed HMXBs in the Milky Way \citep{grimm1}, SMC \citep{smc},
and in the Antennae galaxies \citep{ant}. Note, that the predicted 
$rms-$SFR relation can be modified by the luminosity dependence of the
$rms$ of individual sources. This factor might become especially
important at large values of SFR when the total luminosity of a star
forming galaxy is dominated by ULXs whose variability properties we
know little about.

\begin{figure*}
\centerline{\hbox{
\resizebox{0.45\hsize}{!}{\includegraphics{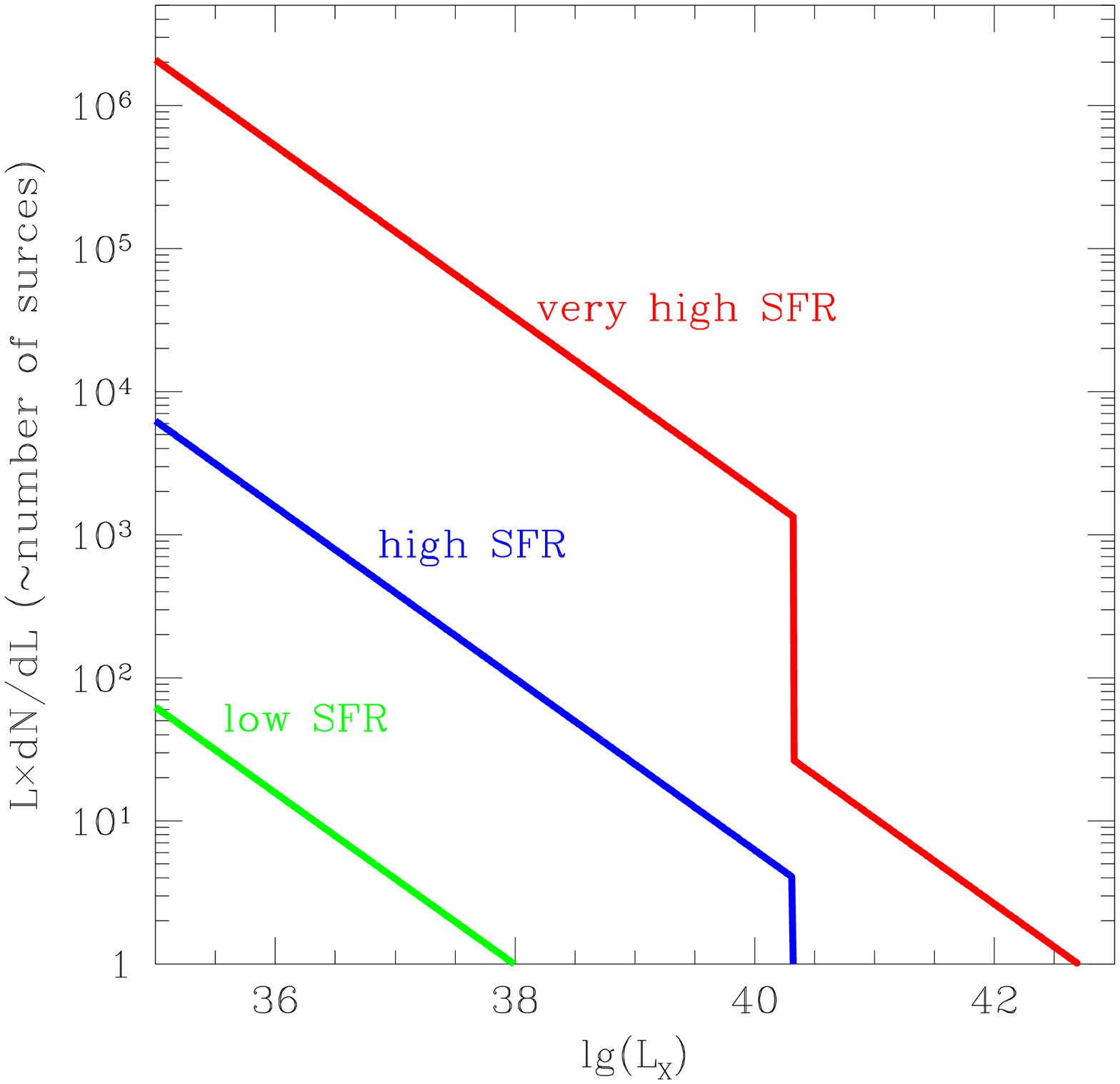}}
\resizebox{0.45\hsize}{!}{\includegraphics{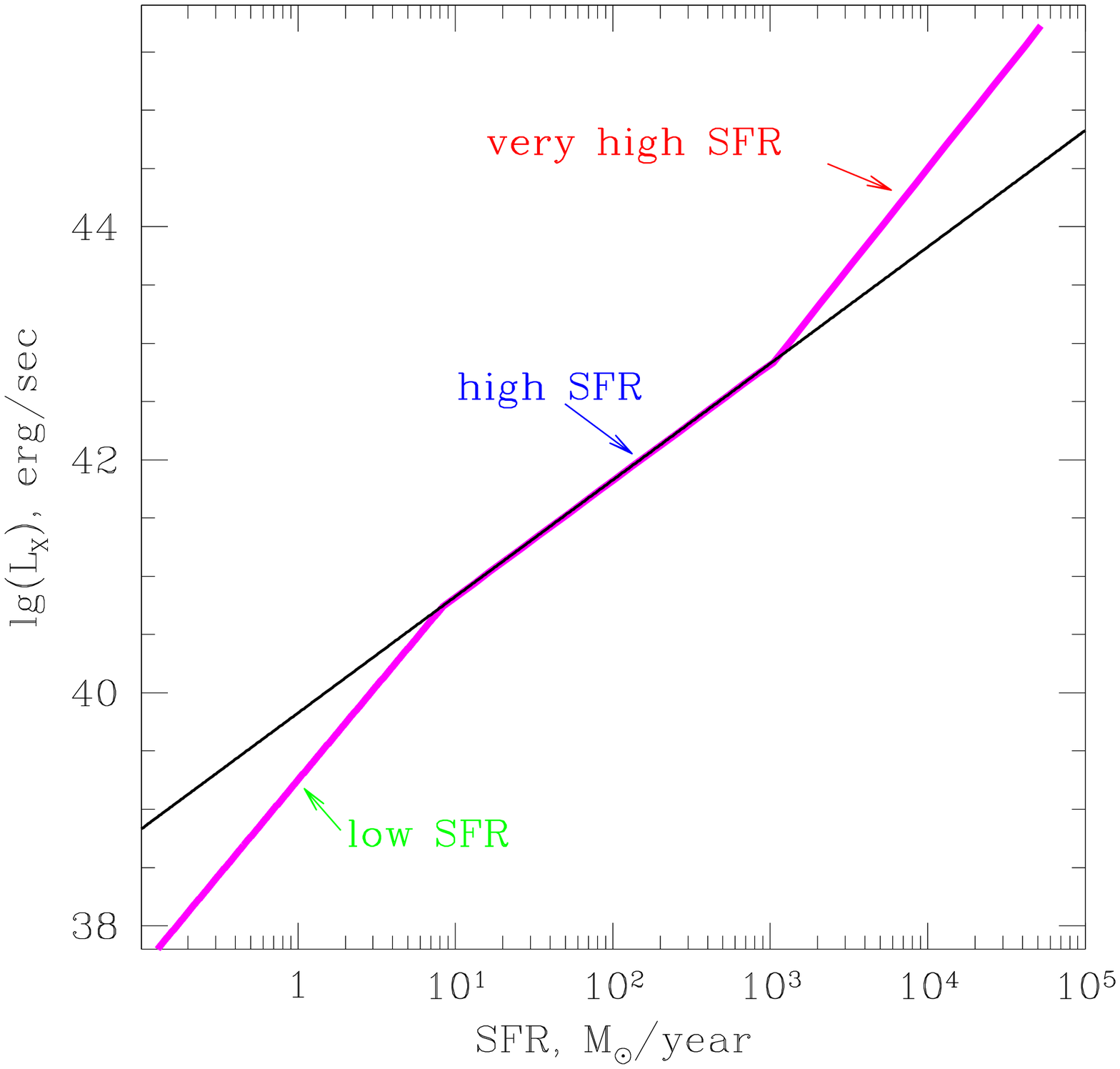}}
}}
\caption{Illustration of the effect of hypothetical intermediate mass
black holes on the $L_X-$SFR relation. 
{\em Left:} The luminosity function of compact sources at different
levels of star formation rate. 
{\em Right:} Corresponding $L_X-$SFR relation. The thin straight line
shows the linear dependence.
\label{fig:imbh}}
\end{figure*}

\subsubsection{Luminosity of the brightest source}

As the first Chandra observations of compact sources in nearby
galaxies became available, it has been noted 
\citep[e.g.][]{ ngc4697, ngc1291,fabbiano2003} that the
luminosity of the brightest  X-ray binary in a galaxy might depend on
its properties. In particular in the case of high mass X-ray binaries
it appeared to correlate with the star formation rate of the host
galaxy. For example, in the Antennae galaxies, a number of compact
sources have been discovered with luminosities of $\sim 10^{40}$
erg/s \citep{ant}. On the other hand, the luminosities of the
brightest HMXB sources in the Milky Way do not exceed $\la 10^{38}$
erg/s \citep{grimm1}. It has been argued that this might reflect the
difference in the intrinsic source properties, related to the
difference in the galactic environment and in initial conditions for
X-ray binary formation in starburst galaxies and in galaxies with
weak and steady star formation. 

However, as discussed in section \ref{sec:prob_lmax}, the probability
distribution for the luminosity of the brightest source in a galaxy,  
$p\,(L_{\rm max})$, depends on the LF normalization, i.e. on the SFR
of the host galaxy in the case of HMXBs. 
The luminosity of the brightest source increases with SFR, until it 
reaches the maximum possible value, defined by the high
luminosity cut-off of the LF. 
The $p\,(L_{\rm max})$  distributions, computed from
eq.(\ref{eq:p_lmax}) for the parameters of the  ``universal'' HMXB
XLF, are shown for different values of SFR in the left panel of
Fig.\ref{fig:lmax_hmxb}. The right panel in Fig.\ref{fig:lmax_hmxb}
shows the dependence of the most probable value of the luminosity of
the brightest HMXB on the SFR of the host galaxy, described by 
eq.(\ref{eq:lmax}), along with its intrinsic 67\% uncertainty. 
Filled symbols in Fig.\ref{fig:lmax_hmxb} are the luminosities of the 
brightest source observed in star forming galaxies from the sample of
\citet[][ and references therein]{grimm}. 
As is clear from the plot, the large difference in
the maximum luminosity  between low- and high-SFR galaxies,
e.g. between the Milky Way and the Antennae galaxies, can
be naturally understood in terms of the properties of the probability
distribution $p\,(L_{\rm max})$. No additional physical
processes, affecting HMXBs formation in starburst galaxies, need to be
invoked.

\subsection{Intermediate mass black holes}
\label{sec:imbh}

The hypothetical intermediate mass black holes, probably reaching
masses of $\sim 10^{2-5} M_{\odot}$, might be produced, e.g. via black 
hole merges in dense stellar clusters, and can be associated with 
extremely high star formation rates. To accrete efficiently they
should form close binary systems with normal stars or be located  in
dense molecular clouds. It is natural to expect, that such objects are
significantly less frequent than $\sim$stellar mass black holes. 
The transition from $\sim$stellar mass BH HMXB to intermediate
mass BHs should manifest itself as a step in the luminosity
distribution of compact sources (Fig.\ref{fig:imbh}, left panel).  
If the cut-off in the HMXB XLF, observed at $\log(L_{\rm cut}) \sim
40.5 $ corresponds to the maximum possible luminosity of ``ordinary''
$\sim$stellar mass black holes and if at $L>L_{\rm cut}$ a population of 
hypothetical intermediate mass BHs emerges, it should lead to
a drastic change in the slope of the $L_X$--SFR relation at extreme
values of SFR  (Fig.\ref{fig:imbh}, right panel). 
Therefore, observations of $L_X-$SFR relation for distant star forming
galaxies with very high SFR might be an easy way to probe the
population of intermediate mass black holes.

\begin{figure*}
\centerline{
\resizebox{0.5\hsize}{!}{\includegraphics{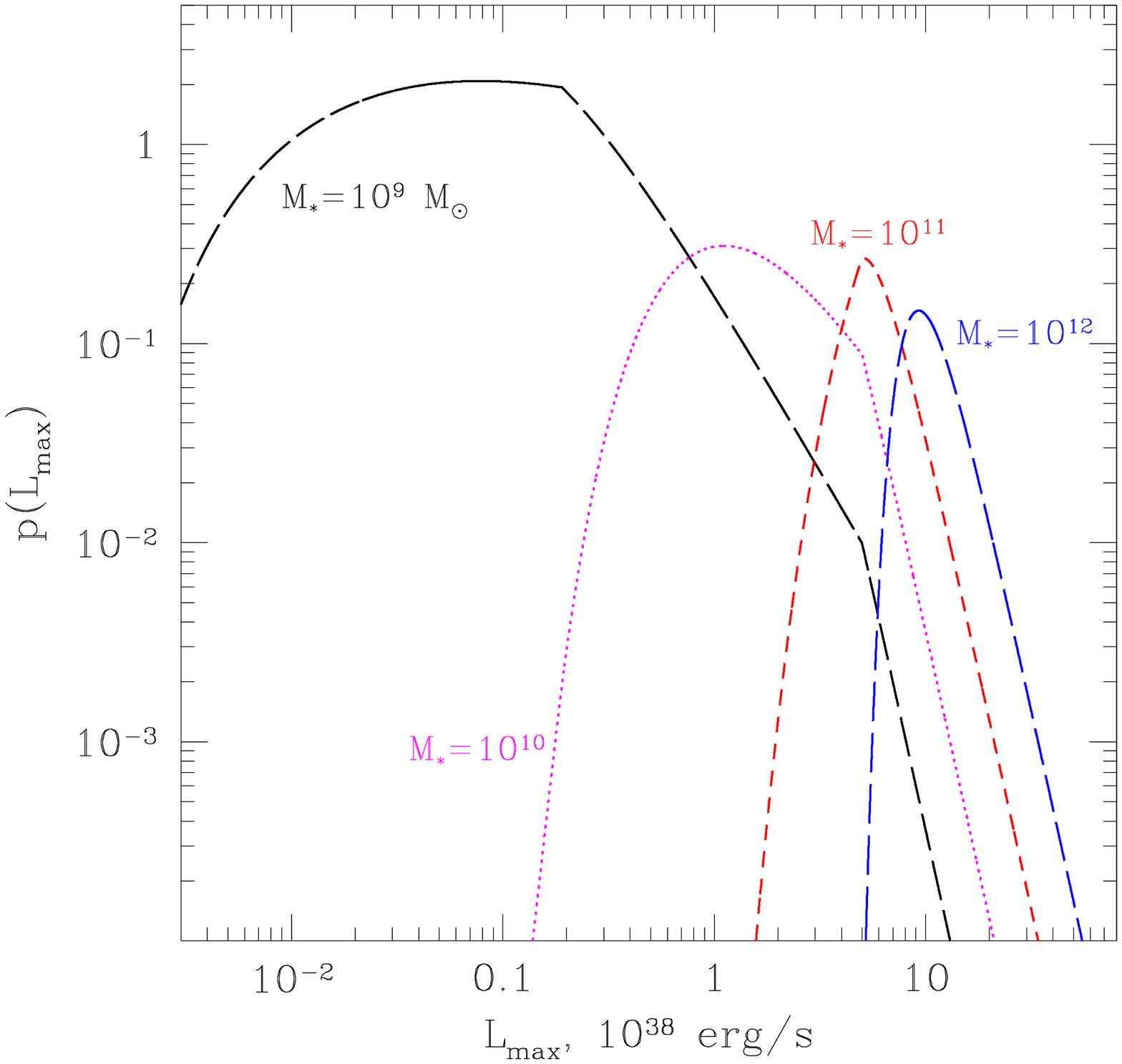}}
\resizebox{0.5\hsize}{!}{\includegraphics{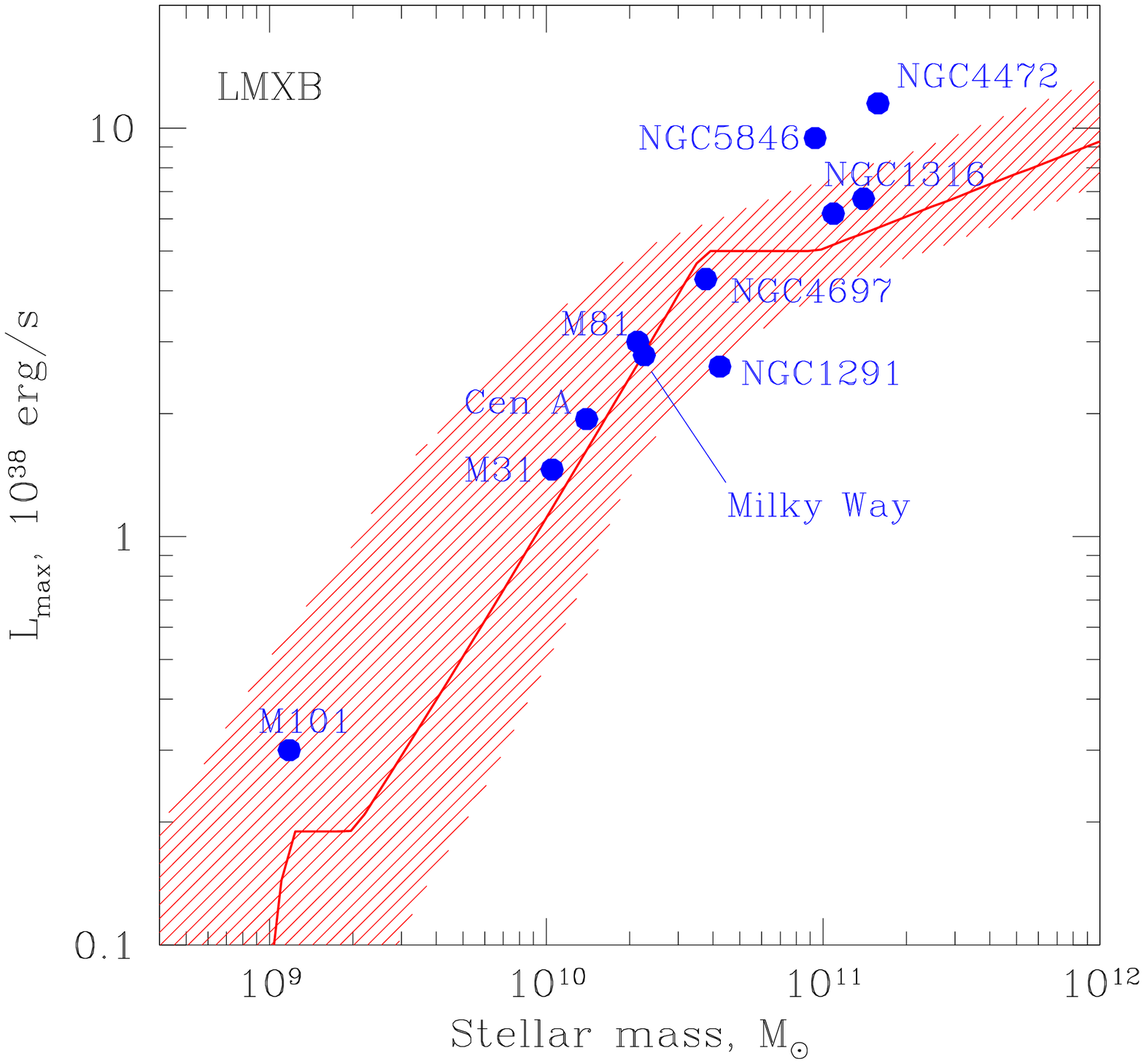}}
}
\caption{{\em Left:} Probability distribution for the luminosity of
the brightest LMXB  for different values of the stellar mass,
computed from eq.(\ref{eq:p_lmax}), using the parameters of the broken
power law approximation to the ``universal'' luminosity function of
LMXBs.    
{\em Right:} Expected luminosity of the brightest LMXB source vs. 
stellar mass of the host galaxy. The thick solid line shows the 
most probable value 
of the $L_{\rm max}$, the shaded area shows its 67\% intrinsic 
dispersion, both obtained from the probability distribution given by
eq.(\ref{eq:p_lmax}). 
The filled circles show maximum observed luminosities of LMXB sources
in the Milky Way and several nearby galaxies, studied with Chandra. 
The ``broken'' shape of the predicted dependence is a consequence of the
broken power law approximation of the ``universal'' LMXB XLF, used in
the calculations.
}
\label{fig:lmax_lmxb}
\end{figure*}

\subsection{Low mass X-ray binaries}
\label{sec:lmxb}

\subsubsection{$L_X-$stellar mass relation and maximum luminosity}

As was shown by \citet{lmxb} the luminosity distribution of low mass
X-ray binaries in nearby early type galaxies and bulges of spiral
galaxies can be described by a ``universal'' XLF whose
shape is approximately the same in different galaxies and whose
normalization is proportional to the stellar mass of a galaxy. The
shape of the ``universal'' LMXB XLF is significantly more complex than
that of HMXBs. It appears to follow the $L^{-1}$ power law at low
luminosities, gradually steepens at $\log(L_X)\ga 37.0-37.5$, and has
a rather abrupt cut-off at $\log(L_X)\sim 39.0-39.5$. In the 
$\log(L_X)\sim 37.5-38.7$ luminosity range, it approximately follows a
power law with the differential slope of $\approx 1.8-1.9$.

Given the shape of the  XLF, the total luminosity of LMXB sources
in a galaxy
is defined by the sources with $\log(L_X)\sim 37-38$, the contribution
of the brighter and, especially, weaker sources being less
significant. Therefore the non-linear regime in the $L_X-M_*$ relation,
although does exist for $\log(M_*)\la 10.0-10.5$, is  less
important than in the $L_X-$SFR relation for high mass X-ray
binaries (see, for example, Fig.14 in \citealt{lmxb}). 

Significantly more pronounced is the dependence of the luminosity of
the brightest source on the LF normalization, i.e. on the stellar
mass of the host galaxy. In order to study this dependence we used the
broken power law approximation for the LMXB XLF from \citet{lmxb} and
performed a series of the Monte-Carlo simulations, similar to those
described in section \ref{sec:variability}. The probability
distribution of the maximum luminosity $p\,(L_{\rm max})$, obtained from 
these simulations, is shown in the left panel in
Fig.\ref{fig:lmax_lmxb} for different values of the stellar mass of
the host galaxy. The right panel shows the dependence of the most
probable value of the maximum luminosity and of its 67\% intrinsic
spread on the stellar mass. Its broken line shape is a result of the 
broken power law approximation of the LMXB 
XLF used in the simulations. Solid symbols show
the observed values of the maximum luminosity for the number of nearby
early type galaxies, bulges of spiral galaxies and for LMXBs in the
Milky Way from the sample of \citet[][ and references therein]{lmxb}.

Similarly to HMXBs in star forming galaxies it is obvious from
Fig.\ref{fig:lmax_lmxb} that the significant difference in the value
of the luminosity of the brightest source can be naturally explained
by the properties of the luminosity function of LMXBs. The same effect 
leads to an artificial (unphysical) dependence of the
average luminosity of low mass X-ray binaries in a galaxy on its
stellar mass (e.g. Fig.17 in \citealt{lmxb}). So far there is no
evidence for a significant change of intrinsic properties of low
mass X-ray binaries with galactic environment. The difference between
the luminosity of the brightest LMXB in massive elliptical galaxies
and the bulges of spiral galaxies can be understood based on the
probability arguments.

\subsubsection{Variability}

\begin{figure}
\centerline{
\resizebox{\hsize}{!}{\includegraphics{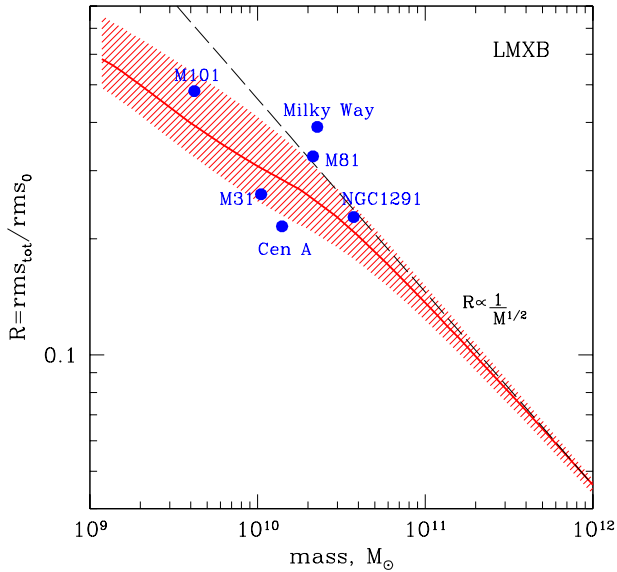}}
}
\caption{Variability of the total emission of low mass X-ray
binaries. Dependence of the ratio 
$rms_{\rm\,tot}/rms_0$ on the stellar mass of the galaxy.
The thick solid line shows the most probable value of the
$rms_{\rm\, tot}/rms_0$, the shaded area shows its 67\% intrinsic
dispersion, both obtained from the Monte-Carlo simulations for the
``universal'' luminosity function from \citet{lmxb}. 
The dashed line shows asymptotical behavior at large $M_*$. The filled
circles correspond to LMXBs in the Milky Way and several  nearby
galaxies, computed from eq.(\ref{eq:rmstot}), using 
the observed luminosities of X-ray sources in these galaxies.
}
\label{fig:rms_lmxb}
\end{figure}

As in the case of HMXBs, in the linear large-mass limit, 
$\log(M_*)\ga 10.5$, the fractional $rms$ of the aperiodic variability
of the combined emission of LMXBs  follows the $rms_{\rm\, tot}\propto
1/\sqrt{M_*}$ averaging law. Owing to the shape of the LMXB
XLF it decreases rather quickly with the stellar mass of the galaxy
in the non-linear low-mass regime as well (Fig.\ref{fig:rms_lmxb}).
Consequently, in massive elliptical galaxies, with
stellar mass $\log(M_*)\sim 11.0-11.5$, the fractional $rms$
variability of the total emission will be suppressed by a factor of
$\sim 10-15$ with respect to the $rms$ of individual sources. 
In a galaxy similar to the Milky Way with $\log(M_*)\sim 10.5-10.7$
the suppression factor is $\sim 5$.
Considerable variability on the level of $\sim 1/4-1/2$ of that of
individual X-ray binaries can be expected only for light bulges of
spiral galaxies with masses in the $\log(M_*)\sim 9.5-10.5$ range.

\begin{figure}
\centerline{
\resizebox{\hsize}{!}{\includegraphics{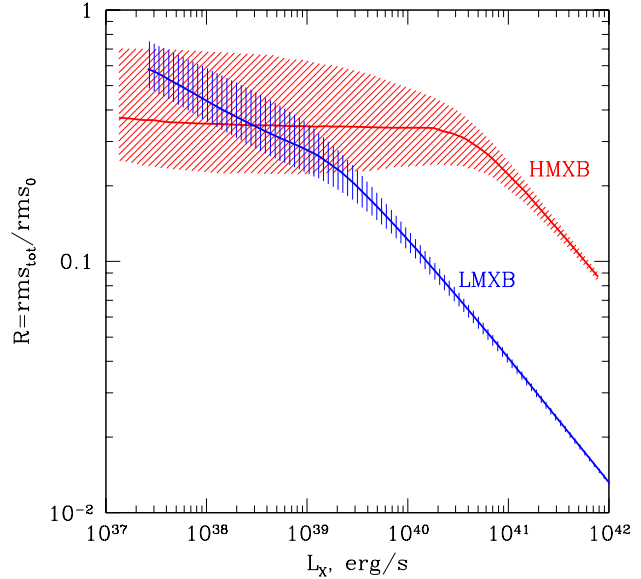}}
}
\caption{Comparison of variability of low and high mass X-ray
binaries. 
Dependence of the ratio $rms_{\rm\,tot}/rms_0$ on the total
X-ray luminosity of the galaxy due to X-ray binaries.
The thick solid lines show the most probable value of the
$rms_{\rm\,tot}/rms_0$, the shaded areas show its 67\% intrinsic
dispersion, both obtained from the Monte-Carlo simulations for the
respective ``universal'' luminosity functions from \citet{grimm} and
\citet{lmxb}.  For both curves, the linear parts at high $L_X$ follow
$rms\propto 1/\sqrt{L_X}$ averaging law.
}
\label{fig:rms_ltot}
\end{figure}

Fig.\ref{fig:rms_ltot} compares the dependence of the fractional $rms$ 
on the most probable value of the total luminosity for high and low
mass X-ray binaries. In the bright luminosity end, $\log(L_X)\ga
39.5$, the X-ray emission from early type galaxies is expected to be
significantly, up to a factor of $\sim 7$, less variable than from
star forming galaxies.

\section{Summary}
\label{sec:summary}

We studied the statistical properties of the combined emission of a
population of discrete sources. Namely, we considered the properties
of their total luminosity
\begin{eqnarray}
L_{\rm tot}=\sum_{k=1}^{k=n} L_k
\end{eqnarray}
and its dependence on the number of sources $n$ or, equivalently, on
the normalization of the luminosity function (LF). Using high mass
X-ray binaries in star forming galaxies as an example, $L_k$ are the
luminosities of individual X-ray binaries in a galaxy and $L_{\rm
tot}$ is its total X-ray luminosity due to HMXB population. In this
example the normalization of the luminosity function, i.e. the number
of HMXBs in the galaxy $n$, is proportional to its star formation rate.
We showed that due to statistical properties of the probability distribution
$p\,(L_{\rm tot})$ the result of a measurement of the total luminosity
of a randomly chosen galaxy might deviate significantly from the
intuitively  obvious expression
\begin{eqnarray}
\left  <L_{\rm tot} \right > =
\int_0^{+\infty} L\,\frac{dN}{dL}\, dL 
\propto n \propto {\rm SFR}
\label{eq:ltot_summary}
\end{eqnarray}
These properties of $p\,(L_{\rm tot})$ can result in a surprising
non-linear dependence of the total luminosity  on $n$. 
They can also cause anomalous variability of the combined emission
in an apparent violation of the $rms\propto 1/\sqrt{n}\,$ averaging
law for uncorrelated variations of individual sources. 

\bigskip

Our results can be summarized as follows:
\begin{enumerate}

\item 
The probability distribution $p\,(L_{\rm tot})$ can be computed
numerically from eq.(\ref{eq:p_ltot}) or alternatively
eq.(\ref{eq:pn_ltot}). Examples are shown in Fig.\ref{fig:probdist}.

\item 
The relevant characteristics of the $p\,(L_{\rm tot})$  distribution
are: (1) its mode $\tilde{L}_{\rm tot}$ -- the value of the random
variable $L_{\rm tot}$ for which $p\,(L_{\rm tot})$ has the maximum
and (2) the expectation mean $\left< L_{\rm tot} \right>$
defined by eq.(\ref{eq:ltot_summary}).

It is the mode of the $p\,(L_{\rm tot})$ distribution that predicts
the most probable value of the total luminosity of a randomly chosen
galaxy. If many galaxies with $\sim$same total number of
sources $n$ are observed the measured values of their total X-ray
luminosities are 
distributed according to the $p\,(L_{\rm tot})$ distribution whose
shape depends on the chosen value of $n$ (Fig.\ref{fig:lx-sfr}, left
panel). The average of the measured values of $L_{\rm tot}$
is equal to the expectation mean $\left< L_{\rm tot} \right>$ and
is always proportional to $n$ in agreement with
eq.(\ref{eq:ltot_summary}).

\item
For small values of the LF normalization, $\tilde{L}_{\rm tot}$ and
$\left< L_{\rm tot} \right>$ do not equal each other
(Fig.\ref{fig:probdist}) and the $\tilde{L}_{\rm tot}-n$ relation is
non-linear (Fig.\ref{fig:lprobn}--\ref{fig:lprob}, \ref{fig:lx-sfr}). 
Only in the limit of $n\gg1$, 
$\tilde{L}_{\rm tot}=\left< L_{\rm tot} \right>$ and  
$\tilde{L}_{\rm tot}$ depends on $n$ linearly.  
The threshold value of $n$ depends on the LF shape and can be
arbitrarily large.

\item The skewness of the $p\,(L_{\rm tot})$ probability distribution
in the non-linear regime (Fig.\ref{fig:probdist}) results in an
enhanced and significantly asymmetric intrinsic dispersion of the
measured values of  $L_{\rm tot}$
(Fig.\ref{fig:lprob},\ref{fig:lx-sfr}).  
Its non-Gaussianity precludes the use of the standard fitting
techniques in analyzing the $L_{\rm tot}-n$ relation (e.g. $L_X-$SFR
relation for star forming galaxies), such as $\chi^2$ minimization
technique.

\item 
The variability of the total emission (e.g. aperiodic variability of
X-ray emission of a galaxy due to superposition of variabilities of
individual sources) in the non-linear regime decreases with the
number of sources more slowly than $rms \propto 1/\sqrt{n}\,$ law for
uncorrelated variations of individual sources, resulting in
anomalously high variability of the total emission. In the linear
regime the $rms \propto 1/\sqrt{n}$ dependence is restored (section
\ref{sec:variability}, Fig.\ref{fig:var}).  

\item The amplitude of the discussed effect depends on the shape of
the luminosity function. For a power law LF it is strongest for
$1<\alpha<2$, is unimportant for shallow luminosity functions with
$\alpha<1$, and gradually diminishes with increase of
$\alpha$ above $\alpha=2$ (Fig.\ref{fig:probdist}--\ref{fig:lprob}).

\end{enumerate}

We illustrate these results using the example of the combined emission 
of X-ray binaries in galaxies and its dependence on the star formation
rate and on the stellar mass of the host galaxy. 

\begin{enumerate}

\item 
For the slope of the HMXB ``universal'' XLF, 
$\alpha\approx 1.6$, the discussed effects are strongest with
a significant non-linear regime in the $L_X-$SFR relation at SFR$\la
4-5$ $M_\odot$/yr. The predicted $L_X-$SFR dependence is in good
agreement with observations (Fig.\ref{fig:lx-sfr}). Given the shape of
the  ``universal'' LMXB XLF no significant non-linearity of the
$L_X-M_*$ relations is expected, also in a good agreement with
observations.

\item
The $L_X-$SFR relation can be used to constrain the XLF parameters of
HMXBs in distant unresolved galaxies including the galaxies observed
with  Chandra in HDF-N at redshifts  $z\sim 0.2-1.3$ (sections
\ref{sec:cutoff_hmxb}, \ref{sec:imbh}, Figs.\ref{fig:lx-sfr-break},
\ref{fig:imbh}).  

\item Both for high and low mass X-ray binaries a strong dependence of
the luminosity of the brightest source on the SFR and
stellar mass of the host galaxy is expected. The  
$L_{\rm max}-$SFR and $L_{\rm max}-M_*$ dependences predicted from
the respective ``universal'' XLFs explain well the results of
Chandra observations of nearby galaxies (Fig.\ref{fig:lmax_hmxb} and
\ref{fig:lmax_lmxb}). The significant difference in the luminosity of
the brightest LMXB 
between bulges of spiral galaxies and giant ellipticals or between the
brightest HMXB in the Milky Way and in starburst galaxies can be
understood based solely on probability arguments.

\item We predict enhanced variability of X-ray emission from star
forming galaxies due to HMXBs, significantly above the 
$\propto 1/\sqrt{n}$ averaging law. For SFR$\la 5$ M$_\odot$/yr the
expected fractional $rms$ of variability of the combined emission of
HMXBs does not depend on the star formation rate and approximately
equals ~$\sim 1/3-1/2$ of the  $rms$ of individual sources
(Fig.\ref{fig:rms_hmxb}). On the contrary, variability of X-ray
emission from early type galaxies due to LMXBs will be significantly
suppressed because of the averaging effect (Fig.\ref{fig:rms_lmxb}),
upto $\sim 3-10$ times in the  $\sim 10^{10}-10^{11}$ M$_\odot$
stellar mass range. For the same total luminosity star forming
galaxies are expected to have significantly larger fractional $rms$,
than massive elliptical and S0 galaxies assuming that fractional
the $rms$ of individual sources are comparable (Fig.\ref{fig:rms_ltot}).

\end{enumerate}

\section{Acknowledgements}

We thank the anonymous referee for the comments, which helped to
improve the presentation of the paper.

{}

\appendix

\section{Approximate solution for the total luminosity}

\label{sec:lprob_approx}

We consider the case of a power law LF, eq.(\ref{eq:lfd_pl}), with a
slope $\alpha>0$. The probability distribution for the luminosity of
the brightest source in the sample of $n$ sources is defined by
eq.(\ref{eq:p_lmax}). The maximum of this distribution gives the most
probable value of the luminosity of the brightest sources:
\begin{eqnarray}
\tilde{L}_{\rm max}=\min\left(L^{\prime}_{\rm max}\,,L_2\right)
\label{eq:lmax}
\\
\nonumber
\left(\frac{L^{\prime}_{\rm max}}{L_1}\right)^{\alpha-1}=
1+\frac{\alpha-1}{\alpha}\,(n-1)
\end{eqnarray}

Similarly, the probability distribution of the minimum luminosity in
the sample is: 
\begin{eqnarray}
p\,(L_{\rm min})= \left[ p_1(L>L_{\rm min}) \right] ^{n-1} 
p_1(L_{\rm min})\ n
\label{eq:p_lmin}
\end{eqnarray}
Contrary to $p\,(L_{\rm max})$ $p\,(L_{\rm min})$ declines steeply at
$L>L_1$ for any $n$ (for $\alpha>0$). Within the accuracy of this
approximation we can assume $p\,(L_{\rm min})=\delta(L_{\rm
min}-L_1)$, i.e. $L_{\rm min}=L_1$.

The total luminosity of $n$ sources distributed between $L_1$ and
$L_{\rm max}$ according to the power law with the slope of $\alpha$
($\alpha\ne 1,\ \alpha\ne 2$) can be approximated as:
\begin{eqnarray}
L_{\rm tot}\approx n\; \frac{1-\alpha}{2-\alpha}\;
\frac{L_{\rm max}^{2-\alpha}-L_1^{2-\alpha}}
{L_{\rm max}^{1-\alpha}-L_1^{1-\alpha}}
\label{eq:ltot_of_lmax}
\end{eqnarray}

Knowing the probability distribution  $p\,(L_{\rm max})$ the
probability distribution $p_n(L_{\rm tot})$ can be calculated as:
\begin{eqnarray}
p_n(L_{\rm tot}) \approx p\,(L_{\rm max})\cdot
\left( \frac{d L_{\rm tot}}{d L_{\rm max}} \right)^{-1}
\label{eq:pn_of_ltot}
\end{eqnarray}
where $L_{\rm max}=L_{\rm max}(L_{\rm tot})$ is inverse function to
eq.(\ref{eq:ltot_of_lmax}) and $p\,(L_{\rm max})$ is given by
eq.(\ref{eq:p_lmax}).

The most probable value of $L_{\rm tot}$ is defined by the condition 
\begin{eqnarray}
\frac{dp_n(L_{\rm tot})}{dL_{\rm tot}}=0
\end{eqnarray}
With eq.(\ref{eq:p_lmax}), (\ref{eq:ltot_of_lmax}) and
(\ref{eq:pn_of_ltot}) the above equation can be transformed to:
\footnote{Note that in eq.(\ref{eq:n_xi_eq}) and
(\ref{eq:n_of_xi}) $L_{\rm max}$ is a parameter
rather than the most probable value of the maximum luminosity. The
latter is defined by the eq.(\ref{eq:lmax}), which is exact.
\label{fn:lmax}}
\begin{eqnarray}
(\alpha-2)\xi^{2\alpha}-
(\alpha-2)(1+\alpha-n+\alpha n)\xi^{1+\alpha}+
\label{eq:n_xi_eq}
\\
+(\alpha-1)^2(1+n)\xi^\alpha+
\left[ 1+(\alpha-1)n\right] \xi ^2 =0
\nonumber
\\
\xi=\frac{L_{\rm max}}{L_1}
\nonumber
\end{eqnarray}
or, equivalently:
\begin{eqnarray}
n=\frac{(\alpha-2)\xi^{2\alpha}+(2+\alpha-\alpha^2)\xi^{1+\alpha}+(\alpha-1)^2\xi^\alpha-\xi^2}
{(\alpha-1)[(\alpha-2)\xi^{1+\alpha}-(\alpha-1)\xi^\alpha+\xi^2]}
\label{eq:n_of_xi}
\end{eqnarray}
$\alpha\ne 1,\ \alpha\ne 2$.
Because of the simplifying assumption 
$p_{\rm min}(L_{\rm min})=\delta(L_{\rm min}-L_1)$
and the approximate nature of eq.(\ref{eq:ltot_of_lmax}),
the probability distribution defined by eq.(\ref{eq:pn_of_ltot}),
is valid only for $L_{\rm tot} < \left< L_{\rm tot} \right>$ and is
undefined otherwise. 
This, however, is sufficient for our purpose as $L_{\rm max}\sim L_2$
corresponds to the break in the $\tilde{L}_{\rm tot} - n$ relation
(Fig.\ref{fig:lprobn} and \ref{fig:lprob}), above which   
$\tilde{L}_{\rm tot}=\left< L_{\rm tot} \right>$. 

\subsection{The practical recipe}
\label{sec:lprob_approx_recipe}

The $\tilde{L}_{\rm tot}-n$ relation can be 
computed parametrically using eqs.(\ref{eq:n_of_xi}) and
(\ref{eq:ltot_of_lmax}) The practical recipe is for a set of values of
$L_{\rm max}$, $L_1 < L_{\rm max} \le L_2$, to compute $n$ from
eq.(\ref{eq:n_of_xi}) and $L_{\rm tot}$ from
eq.(\ref{eq:ltot_of_lmax}). The pairs of values 
$(L_{\rm tot},n)$ define the $\tilde{L}_{\rm tot} - n$ relation
before and up to the break. Above the break, 
$\tilde{L}_{\rm tot}=\left< L_{\rm tot} \right>$ and can be computed
from eq.(\ref{eq:ltot_of_lmax}) with $L_{\rm max}=L_2$ and 
$n>n_{\rm break}$ -- free parameter. The $n$ in the obtained
$\tilde{L}_{\rm tot} - n$ relation can be transformed to the
normalization $A$ via eq.(\ref{eq:ntot}).

The approximation defined by the eqs.(\ref{eq:n_of_xi}) and
(\ref{eq:ltot_of_lmax}) is compared with the results of the exact
calculation in Fig.\ref{fig:lprobn}. It is accurate within $\sim$
several per cent everywhere, except the break region, where its
accuracy is $\sim 10-20\%$.

\subsection{Asymptotics}
\label{sec:asympt}

Using the approximate solution for $\tilde{L}_{\rm tot}$ from Appendix 
\ref{sec:lprob_approx} we consider the asymptotical behavior of the
$\tilde{L}_{\rm tot} - n$ relation in the limit of $L_2/L_1\rightarrow
\infty$, $\alpha>1$.

From eq.(\ref{eq:n_of_xi}) variable $\xi$ is related to the number of
sources by (see footnote \ref{fn:lmax}): 
\begin{eqnarray}
\xi^{\alpha-1}=(\alpha-1)n+O(1)
\nonumber
\\
L_{\rm max}=\xi L_1
\label{eq:ksi_of_n_asympt}
\end{eqnarray}
in the limit of $n>>1$ or equivalently $L_{\rm max}>>L_1$. Although in
the limit $n\rightarrow\infty$ this approximation is valid for any
$\alpha>1$, its accuracy deteriorates considerably for $\alpha\la
1.6$ where it can be improved by replacing $n$ in
eq.(\ref{eq:ksi_of_n_asympt}): 
\begin{eqnarray}
n\rightarrow n-\frac{1+\alpha-\alpha^2}{(\alpha-1)(\alpha-2)},
~~ 1<\alpha\la 1.6
\label{eq:n_replacement}
\end{eqnarray}
The most probable value of the total luminosity is given by
\begin{eqnarray}
\frac{\tilde{L}_{\rm tot}}{\left< L_{\rm tot}\right>}\approx 
\frac{
1-\left[ (\alpha -1) \, n\,\right]^{\frac{2-\alpha}{\alpha -1}}}
{1-\left(L_2/L_1\right)^{2-\alpha}}
\label{eq:ltot_vs_n_asympt}
\end{eqnarray}
$\alpha>1$, $\alpha\ne 2$, $n>>1$.
Using eq.(\ref{eq:ntot}), (\ref{eq:ltot}) it can be expressed via the
normalization of the luminosity function $A$ or transformed to the
relations for $\tilde{L}_{\rm tot}$. 
As with eq.(\ref{eq:ksi_of_n_asympt}), the accuracy of 
eq.(\ref{eq:ltot_vs_n_asympt})  can be significantly improved using
eq.(\ref{eq:n_replacement}) for $\alpha\la 1.6$.   
 
For $1<\alpha<2$ the $\tilde{L}_{\rm tot}-n$ relation shows a 
sharp break between the non-linear and linear regimes
(Fig.\ref{fig:lprobn},\ref{fig:lprob}). 
From eq.(\ref{eq:ltot_vs_n_asympt}) one can obtain:
\begin{eqnarray}
\renewcommand{\arraystretch}{1.5}
\tilde{L}_{\rm tot}\propto 
\left\{ \begin{array}{ll}
n^{\frac{1}{\alpha-1}} & n<n_{\rm break}\\
n & n>n_{\rm break}
\end{array} \right.
\label{eq:ltot_small_n_pl}
\end{eqnarray}
This is in an agreement with eq.(\ref{eq:ltot_intro}) based on simple
qualitative arguments. The position of the break in the
$\tilde{L}_{\rm tot} - n$ relation can be obtained from
eq.(\ref{eq:n_of_xi}) substituting $\xi=\frac{L_2}{L_1}$ and using the
fact that $\xi>>1$:
\begin{eqnarray}
n_{\rm break}\approx \frac{1}{\alpha-1} \cdot 
\left( \frac{L_2}{L_1} \right )^{\alpha-1}
\label{eq:nbreak}
\end{eqnarray}
Expressed in terms of the the normalization $A$ of the luminosity
function eq.(\ref{eq:lfd_pl}) it is:
\begin{eqnarray}
A_{\rm break}\approx L_2^{\alpha-1}
\label{eq:abreak}
\end{eqnarray}
As intuitively expected, the break position expressed in terms of the
normalization of the luminosity function does not depend on the low
luminosity cut-off $L_1$ and is defined only by the slope of the
luminosity function and the high luminosity cut-off (see discussion in
section \ref{sec:ltot}). 
The total luminosity at the break, however, depends on the low
luminosity cut-off for steep luminosity function with $\alpha>2$:
\begin{eqnarray}
\renewcommand{\arraystretch}{1.5}
L_{\rm tot,break}\approx
\left\{ \begin{array}{ll}
\frac{L_2}{2-\alpha}	& \mbox{if $1<\alpha<2$}\\
\frac{L_2}{\alpha-2}\times\left(\frac{L_2}{L_1}\right)^{\alpha-2}&
\mbox{if $\alpha>2$} 
\end{array} \right.
\end{eqnarray}
as the total luminosity for $\alpha>2$ is
defined by the sources near the low luminosity cut-off.

\section{Variability of the total emission}
\label{sec:variability_ap}

In the linear regime of $L_{\rm tot}-n$ relation,  the fractional
$rms^2$ of the collective emission is inversely proportional to the
number of sources:  
\begin{eqnarray}
\renewcommand{\arraystretch}{1.9}
\frac{rms_{\,\rm tot}^2}{rms_0^2}\approx
\left\{ \begin{array}{ll}
\frac{(2-\alpha)^2}{(1-\alpha)(3-\alpha)} \ \frac{1}{n} & \mbox{if $\alpha<1$}\\
\frac{(2-\alpha)^2}{(\alpha-1)(3-\alpha)} \ \left(\frac{L_2}{L_1} \right)^{\alpha-1}
\frac{1}{n} & \mbox{if $1<\alpha<2$}\\
\frac{(2-\alpha)^2}{(\alpha-1)(3-\alpha)} \ \left(\frac{L_2}{L_1} \right)^{3-\alpha}
\frac{1}{n} & \mbox{if $2<\alpha<3$}\\
\frac{(2-\alpha)^2}{(\alpha-1)(\alpha-3)} \ \frac{1}{n} & \mbox{if $\alpha>3$}
\end{array} \right.
\label{eq:rmstot_of_n_large_n}
\end{eqnarray}
or, equivalently, to their total luminosity:
\begin{eqnarray}
\renewcommand{\arraystretch}{1.5}
\frac{rms_{\,\rm tot}^2}{rms_0^2}\approx
\left\{ \begin{array}{ll}
\frac{2-\alpha}{3-\alpha} \ \frac{L_2}{L_{\rm tot}} & \mbox{if $\alpha<2$}\\
\frac{\alpha-2}{3-\alpha} \ \frac{L_2}{L_{\rm tot}} \ 
\left( \frac{L_1}{L_2} \right)^{\alpha-2} & \mbox{if $2<\alpha<3$}\\
\frac{\alpha-2}{\alpha-3} \ \frac{L_1}{L_{\rm tot}} & \mbox{if $\alpha>3$}
\end{array} \right.
\label{eq:rmstot_of_ltot_large_n}
\end{eqnarray}
The above formulae are valid in the limit $L_1<<L_2$.

Similarly to the most probable value of the total
luminosity (Appendix \ref{sec:lprob_approx}) the fractional $rms$ of the
total emission can be approximately calculated substituting $L_2$ in 
eq.(\ref{eq:rmstot_large_n}) with some value $L_{\rm max}\leq L_2$. In
principle, the probability distribution for the $rms_{\rm \, tot}$ in  
eq.(\ref{eq:rmstot}) could be derived using the
probability distribution for $L_{\rm max}$. The maximum of this
probability distribution would give a sufficiently accurate
approximation for $rms_{\,\rm tot}/rms_0$.
However, for simplicity we use the value of $L_{\rm max}$ from 
eq.(\ref{eq:n_of_xi}). In the limit $L_2>>L_1$ one finds
for $1<\alpha<3$: 
\begin{eqnarray}
\frac{rms_{\,\rm tot}^2}{rms_0^2}=
\frac{(\alpha -2)^2}{(\alpha-1)(3-\alpha)} \,
\frac{\xi^{3-\alpha}}{(\xi^{2-\alpha}-1)^2} \, 
\frac{1}{n}
\label{eq:rmstot_of_n_small_n}
\end{eqnarray}
for $n<n_{\rm break}$, where $\xi$ is defined by
eq.(\ref{eq:ksi_of_n_asympt}) with substitution of 
eq.(\ref{eq:n_replacement}) for $1<\alpha\la 1.6$ and 
$n_{\rm break}$ is defined by eq.(\ref{eq:nbreak}).
For $\alpha<1$ and $\alpha>3$ there is no non-linear regime and the
fractional $rms_{\rm tot}$ obeys eq.(\ref{eq:rmstot_of_n_large_n}) 
and (\ref{eq:rmstot_of_ltot_large_n}) for any $n$.

From eq.(\ref{eq:rmstot_of_n_small_n}) one finds for $1<\alpha<2$
\begin{eqnarray}
\frac{rms_{\,\rm tot}^2}{rms_0^2}=
\frac{(\alpha -2)^2}{(3-\alpha)} 
\ \ \ \ \ \ \ \ \ \ n<n_{\rm break}
\label{eq:rmstot_of_n_small_n_1}
\end{eqnarray}
i.e. in the non-linear regime the fractional $rms$ of the
collective emission does not depend on the number of sources.

The accuracy of eqs.(\ref{eq:rmstot_of_n_small_n}) 
and (\ref{eq:rmstot_of_n_small_n_1}) is sufficiently good for
$\alpha\ga 1.5$ but deteriorates for smaller values of $\alpha$. 
In the linear regime eqs.(\ref{eq:rmstot_of_n_large_n}) and
(\ref{eq:rmstot_of_ltot_large_n}) are almost precise, their only
approximation is in neglecting higher orders of the $L_1/L_2$.

\end{document}